\documentclass[10pt,aps,prb,secnumarabic,superscriptaddress,nopacs,amsmath,amstex,amssymb,citeautoscript,longbibliography,floatfix,twocolumn]{revtex4-2}
\usepackage{graphicx,amsmath,amssymb,amsfonts,color}
\usepackage{graphicx}
\usepackage[percent]{overpic}
\usepackage{nicefrac}
\usepackage{multirow, makecell, ragged2e}
\usepackage{physics}
\usepackage{bm}
\usepackage{float,lipsum}
\usepackage{algpseudocode}
\usepackage{algorithm}
\usepackage{placeins}
\usepackage[inline]{enumitem}
\usepackage{svg}
\usepackage[colorlinks,bookmarks=true,citecolor=blue,linkcolor=red,urlcolor=blue]{hyperref}
\usepackage{graphicx}
\usepackage{overpic}
\usepackage[nameinlink,noabbrev]{cleveref}
\usepackage{tikz}


\usepackage{adjustbox}
\usepackage{bm}
\usepackage{color}
\usepackage{braket}
\usepackage{standalone}
\usepackage{multirow}
\usepackage{tikz}
\usepackage{mathrsfs}
\usepackage{dsfont}
\usepackage{comment}
\usepackage{amsmath}
\usepackage{lipsum}
\usepackage{orcidlink}
\usepackage{subcaption}

\usepackage{graphicx,amsmath,amssymb,amsfonts, color}
\usepackage{nicefrac}
\usepackage{multirow, makecell, ragged2e}
\usepackage{bm}
\usepackage{float,lipsum}
\usepackage{algpseudocode}
\usepackage{algorithm}

\newcommand{\be}{\begin{equation}}
\newcommand{\ee}{\end{equation}}

\begin{document}
\raggedbottom
\title{Phase diagram and lattice structure of interlayer vortices in multilayer Ising superconductors in parallel magnetic fields}
\author{Subhradeep Mahata}
\affiliation{Department of Physics, 104 Davey Lab, The Pennsylvania State University, University Park, Pennsylvania 16802, USA}
\author{Uddalok Nag}
\affiliation{Department of Physics, 104 Davey Lab, The Pennsylvania State University, University Park, Pennsylvania 16802, USA}
\author{J. K. Jain\orcidlink{0000-0003-0082-5881}}
\affiliation{Department of Physics, 104 Davey Lab, The Pennsylvania State University, University Park, Pennsylvania 16802, USA}
\affiliation{Center for Theory of Emergent Quantum Matter, The Pennsylvania State University, University Park, Pennsylvania 16802, USA}
\affiliation{Lodha Theoretical Physics Institute, 17th Floor, Lodha NCP Supremus, Wadala (E), Mumbai 400037, India}

\begin{abstract}
In this work, we investigate layered $s$-wave Ising superconductors under an in-plane magnetic field by finding self-consistent solutions of the microscopic Bogoliubov–de Gennes equations. We determine how the superconducting phase diagram and interlayer vortex structure evolve with increasing layer number by comparing bilayer and four-layer systems with the infinite-layer limit. Compared to standard Ginzburg--Landau treatments of this
problem, our approach applies over a broader parameter regime
and self-consistently resolves the spatial and layer dependence
of the superconducting gap amplitude without imposing a
prescribed form. We find that the range of magnetic fields supporting the conventional Bardeen–Cooper–Schrieffer phase (defined as the phase with no interlayer vortices) is reduced with increasing layer number and is essentially absent at nonzero magnetic flux in the infinite-layer system, consistent with recent thickness-dependent experiments on multilayer $\mathrm{2H}\text{-}\mathrm{NbSe}_2$. In the infinite-layer limit, we find that the Fulde–Ferrell–Larkin–Ovchinnikov phase containing interlayer vortices persists to progressively larger in-plane fields as the temperature is lowered below the zero-field critical temperature, consistent with the experimentally observed enhancement of the upper critical field at low temperatures. The vortex structure also evolves with interlayer hopping: in the four-layer system, for the fluxes and system sizes considered, increasing hopping reorganizes the vortex configuration from outer to inner interlayer junctions. At larger interlayer hopping, we find that the gap distortion associated with an interior vortex extends to the surface layers, producing a local suppression of the surface-layer gap amplitude comparable in magnitude to that in the interior layers. Representative six-layer calculations show that the interior vortices likewise produce a finite, although weaker, modulation of the surface-layer gap. These results highlight the layer number and the interlayer tunneling as key control parameters governing the phase diagram and vortex arrangement in multilayer Ising superconductors.
\end{abstract}
\graphicspath{{./Figures/}}

\maketitle

\section {Introduction}\label{sec:intro}The interplay between superconductivity and magnetic fields has given rise to much interesting physics. Magnetic fields can strongly modify a superconducting state, for example by producing vortices~\cite{Abrikosov1957,Tinkham2004} and, under suitable conditions, stabilizing finite-momentum superconducting phases~\cite{Fulde1964,Larkin1965,Matsuda2007,Croitoru2017}. In addition, magnetic fields and spin-orbit coupling (SOC) can provide routes to topological superconductivity~\cite{Sato2017,Flensberg2021, Jeon2019}. Topological superconductors can host Majorana zero modes with non-Abelian exchange statistics~\cite{Nayak2008,Read2000,Kitaev2001,Beenakker2013}, providing a possible platform for fault-tolerant quantum computation~\cite{Nayak2008,Flensberg2021}. These developments motivate further study of how magnetic fields reshape superconducting states and give rise to new emergent phases.

For a system preserving inversion and time-reversal symmetries (TRS), the superconducting order parameter in the Bardeen–Cooper–Schrieffer (BCS) phase is spatially uniform, with electrons of opposite momenta forming Cooper pairs with zero center-of-mass momentum~\cite{Bardeen1957}. However, breaking either of these symmetries,
for example TRS by applying a magnetic field, can allow finite-momentum
Cooper pairs. Such finite-momentum paired states are referred to as Fulde–Ferrell–Larkin–Ovchinnikov (FFLO) superconducting states, with the physical mechanism selecting the finite momentum depending on the system. In the conventional FFLO
state, magnetic-field-induced Zeeman splitting stabilizes Cooper
pairs with a center-of-mass momentum $\mathbf{q}\neq0$, leading to a
spatially varying complex order parameter
$\Delta(\mathbf{r})=|\Delta(\mathbf{r})|
e^{i\mathbf{q}\cdot\mathbf{r}}$~\cite{Fulde1964,Larkin1965}. In
layered Ising superconductors under an in-plane magnetic field,
spin-induced pair breaking is strongly suppressed~\cite{Xi2016}, and
the finite pair momentum instead arises primarily from the
layer-dependent orbital coupling to the magnetic
field~\cite{Liu2017,Nag2025}, giving rise to the orbital-FFLO phase. In this phase, competition between the in-plane flux and interlayer tunneling can lead to interlayer Josephson vortices, which are accompanied by a local suppression of $|\Delta(\mathbf{r})|$ in the adjacent layers. Candidate FFLO phases driven by orbital and Zeeman effects have been reported in transition metal dichalcogenides (TMDs)~\cite{Wan2023,Cao2026} and strongly Pauli-limited superconductors such as CeCoIn$_5$~\cite{Bianchi2003,Matsuda2007}, respectively, although identifying unambiguous FFLO signatures remains challenging~\cite{Croitoru2017}.

The orbital-FFLO state in layered Ising superconductors has been investigated in both theoretical and experimental contexts. A theoretical study by Liu~\cite{Liu2017}
established the orbital-FFLO mechanism in bilayer systems using the
Ginzburg--Landau (GL) formalism, which was subsequently studied by Nag et. al.~\cite{Nag2025} using the microscopic Bogoliubov--de Gennes (BdG)
approach. Finite-layer GL calculations
have examined its evolution with increasing layer
number~\cite{Yuan2023,Yan2026}. Experimentally, the strong protection of Ising superconductivity against in-plane magnetic fields was first demonstrated in monolayer NbSe$_2$~\cite{Xi2016, Xing2017}. Subsequent measurements have reported orbital-effect-driven finite-momentum pairing in centrosymmetric bilayer systems~\cite{Zhao2023}, as well as signatures of orbital-FFLO superconductivity and a pronounced thickness dependence in multilayer $\mathrm{2H}\text{-}\mathrm{NbSe}_2$~\cite{Wan2023,Cao2026}. Scanning
tunneling microscopy (STM) measurements have also established the ability to probe
vortex-induced modifications of the surface local density of states (LDOS)
in $\mathrm{2H}\text{-}\mathrm{NbSe}_2$~\cite{Hess1989,Hess1994,Galvis2018, Fridman2011}. Additionally, recent thermodynamic measurements have likewise reported signatures consistent with an orbital-FFLO state persisting into the bulk limit of $\mathrm{2H}\text{-}\mathrm{NbSe}_2$~\cite{Cho2025BulkFFLO}. These
developments motivate a microscopic, layer-resolved investigation of
how the superconducting phases and interlayer vortex configurations
evolve with increasing layer number.

To address this problem, we obtain self-consistent solutions of the microscopic BdG
equations to investigate multilayer and infinite-layer $s$-wave
Ising superconductors in an in-plane magnetic field. Our results show that increasing the number of layers reduces the stability of the uniform BCS state, while in the infinite-layer system the orbital-FFLO phase remains stable to increasingly large in-plane fields as the temperature is lowered. We further find that increasing interlayer hopping in a four-layer system drives a rearrangement of the Josephson vortices from configurations associated with the outer layers toward configurations localized within the bulk.
Once located in the interior, the associated
deformation of the order parameter extends to
the surface layers, producing a comparable modulation of the
surface gap that may be accessible to surface-sensitive
probes such as STM.

The remainder of this paper is organized as follows. We first introduce the model Hamiltonian and describe the numerical method in Sec.~\ref{sec:model}. We then examine the Josephson-vortex lattice in the infinite-layer limit (Sec.~\ref{sec:infinite}), present the superconducting phase diagrams for different layer numbers in Sec.~\ref{sec:pdiag}, and investigate the interlayer-hopping-driven vortex rearrangement and surface signature of interior vortices in four- and six-layer systems (Sec.\ref{sec:rearrangement}). Finally, we summarize our findings and discuss their experimental implications in Sec.~\ref{sec:discussion}. All relevant additional calculations and numerical consistency checks are presented in the appendices.

{\section{Model Hamiltonian and numerical method}\label{sec:model}}Each superconducting layer is modeled as a square lattice with an
on-site attractive Hubbard interaction. We set the intralayer
hopping to unity, while $t$ denotes the interlayer
hopping. Here, $\sigma\in\{\uparrow,\downarrow\}$ denotes the
electron spin, $l$ labels the superconducting layers, and
$\mathbf{j}=(j_x,j_y)$ denotes the in-plane lattice sites, with
$\hat{\delta}\in\{\hat{\mathbf{x}},\hat{\mathbf{y}}\}$.
The Hubbard-attraction strength is $U>0$ and $\mu$ is the chemical
potential. The Hamiltonian is given by $H=H_0+H_I$, with

\begin{align}
H_{0}
&=-
\sum_{l,\mathbf{j},\hat{\delta},\sigma}
\left(
e^{iA^{l}_{\hat{\delta},\mathbf{j}}}
c^{\dagger}_{l,\mathbf{j}+\hat{\delta},\sigma}
c_{l,\mathbf{j},\sigma}
+\mathrm{H.c.}
\right)
\nonumber\\
&\quad
-t
\sum_{l,\mathbf{j},\sigma}
\left(
e^{iA^{l}_{z,\mathbf{j}}}
c^{\dagger}_{l+1,\mathbf{j},\sigma}
c_{l,\mathbf{j},\sigma}
+\mathrm{H.c.}
\right)
\nonumber\\
&\quad
-\mu
\sum_{l,\mathbf{j},\sigma}
c^{\dagger}_{l,\mathbf{j},\sigma}
c_{l,\mathbf{j},\sigma},
\label{eq:hamiltonian}
\\
H_{I}
&=
-U\sum_{\mathbf{j},l}
c^{\dagger}_{l,\mathbf{j},\uparrow}
c^{\dagger}_{l,\mathbf{j},\downarrow}
c_{l,\mathbf{j},\downarrow}
c_{l,\mathbf{j},\uparrow}.
\end{align}
For a finite-layer system, we impose open boundary conditions along $\mathbf{\hat{z}}$, implying that the interlayer-hopping sum in $(1)$ runs over
$l=1,\ldots,P-1$. The in-plane magnetic field $\textbf{B}=B\hat{y}$ is incorporated through the Peierls phases $A^{l}_{\hat{\delta},\mathbf{j}} $ and $A^l_{z,\mathbf{j}}$, the sum of which around a plaquette is the Aharonov-Bohm phase $2\pi\Phi_p/\Phi_0$ where $\Phi_p$ denotes the magnetic flux passing through the plaquette and $\Phi_0=h/e$ is the flux quantum. For a finite stack of $P$ layers, we choose the inversion-symmetric
gauge
\[
A^{l}_{\hat{x},\mathbf{j}}
=
\pi(P+1-2l)\frac{\Phi_p}{\Phi_0},
\qquad
A^{l}_{\hat{y},\mathbf{j}}=A^l_{z,\mathbf{j}}=0
\]
where $l=\{1,\ldots,P\}$ labels the layers from top to bottom. For finite-layer systems, the in-plane magnetic field does not break translational symmetry, so no magnetic unit cells (MUCs) are required and \(\Phi_p\) can take arbitrary values. In the infinite-layer system, however, the magnetic field breaks translational symmetry of the Hamiltonian, and we therefore use MUCs of different sizes, each enclosing an appropriate integer number of flux quanta. Note that we have neglected the Zeeman coupling term in the Hamiltonian, since the electron spins are locked in the out-of-plane direction in an Ising superconductor. We also omit an Ising SOC term in the Hamiltonian, since its effects can be absorbed into a renormalization of
the interlayer hopping within
the low-energy effective description of TMDs considered here (see Ref.~\cite{Nag2025}).

To construct the mean-field Hamiltonian $H_{MF}$, we replace the interaction term $H_{I}$ in (2) by its mean-field representation:
\begin{align}
H_I \rightarrow
-\sum_{l,\mathbf{j}}
\left(
\Delta_{l,\mathbf{j}}\,
c^{\dagger}_{l,\mathbf{j},\uparrow}
c^{\dagger}_{l,\mathbf{j},\downarrow}
+
\Delta^{*}_{l,\mathbf{j}}\,
c_{l,\mathbf{j},\downarrow}
c_{l,\mathbf{j},\uparrow}
-
\frac{\left|\Delta_{l,\mathbf{j}}\right|^{2}}{U}
\right),
\end{align}
where the superconducting order parameter is defined as
\begin{equation}
    \Delta_{l,\mathbf{j}}
    =
    U\left\langle
    c_{l,\mathbf{j},\downarrow}
    c_{l,\mathbf{j},\uparrow}
    \right\rangle ,
    \label{eq:order_parameter}
\end{equation}
with $\langle\cdots\rangle$ denoting the thermal average. We solve this multilayer Hamiltonian for an $N\times N$ system in the $xy$ plane using antiperiodic boundary conditions (APBCs) in both $\mathbf{\hat{x}}$ and $\mathbf{\hat{y}}$ directions (see also Appendix~\ref{app:bc}).

In finite-layer systems, we take a unit cell spanning the entire
system along $\hat{\mathbf{x}}$ and $\hat{\mathbf{z}}$ and with one
lattice site along $\mathbf{\hat{y}}$. This amounts to $N\times1\times P$ sites in the unit cell for a P-layer system. In the infinite-layer case, we use a four-layer MUC along $\hat{\mathbf{z}}$ to accommodate staggered vortex configurations while keeping the computational cost manageable, with one lattice site along $\hat{\mathbf{y}}$. The extent in the $\mathbf{\hat{x}}$ direction and the number of flux quanta passing through the MUC were chosen appropriately to obtain the various required $\Phi_p$ values for the phase diagram.

To express the Hamiltonian in momentum space, we consider a
unit cell of size $M\times 1\times P$. A lattice site is written as
\begin{equation}
\mathbf{r}_{\mathbf{R},\alpha,l}
=
\mathbf{R}
+
(\alpha-1)\hat{\mathbf{x}}
+
(l-1)\hat{\mathbf{z}},
\end{equation}
where
\begin{equation}
\mathbf{R}
=
mM\hat{\mathbf{x}}
+
j_y\hat{\mathbf{y}}
+
nP\hat{\mathbf{z}}
\end{equation}
labels the unit cells. Here, the indices
$m\in\{0,1,\ldots,N/M-1\}$, $j_y\in\{0,1,\ldots,N-1\}$, and $n\in\mathbb{Z}$
label the unit cells along $\hat{\mathbf{x}}$, $\hat{\mathbf{y}}$ and $\hat{\mathbf{z}}$ respectively, while
$\alpha=1,2,\ldots,M$ and $l=1,2,\ldots,P$ label the
sites and layers within each unit cell, respectively. Here we assume that N is even and divisible by M.
We then use the Fourier transformation
\begin{equation}
c_{\mathbf{R},\alpha,l,\sigma}
=
\frac{1}{\sqrt{N_{\mathrm{uc}}}}
\sum_{\mathbf{k}}
e^{i\mathbf{k}\cdot\mathbf{R}}\,
c_{\alpha,l,\sigma}(\mathbf{k}),
\end{equation}
where $N_{\mathrm{uc}}$ is the total number of unit cells.
The momentum is written as
\begin{equation}
\mathbf{k}
=
\frac{\pi}{N}
\left(
\nu_x\hat{\mathbf{x}}
+
\nu_y\hat{\mathbf{y}}
\right)
+
k_z\hat{\mathbf{z}},
\end{equation}
with
\begin{align*}
\nu_x &\in 2\mathbb{Z}+1,
&
-\frac{N}{M}<\nu_x\leq\frac{N}{M},
\\
\nu_y &\in 2\mathbb{Z}+1,
&
-N<\nu_y\leq N,
\end{align*}
corresponding to APBCs along
$\hat{\mathbf{x}}$ and $\hat{\mathbf{y}}$. Along the stacking
direction, the momentum lies in the reduced Brillouin zone
\begin{equation}
-\frac{\pi}{P}<k_z\leq\frac{\pi}{P}.
\end{equation}
For the \(P=4\) MUC used in the infinite-layer calculations, we fix the system size along $\mathbf{\hat{z}}$ to 64 layers and sample the corresponding \(k_z\) interval using 16 uniformly spaced points (see also Appendix~\ref{app:kz}). For a finite $P$-layer system, we set $M=N$ and omit the
unit-cell index $n$ and momentum $k_z$. The rest of the calculation is analogous for both finite and infinite-layer systems.

For the $M\times 1\times P$ unit cell, we define the $4MP$-dimensional Nambu basis \begin{equation*}
\Psi(\mathbf{k})
=
\left[
\{c_{l,\alpha,\uparrow}(\mathbf{k})\},
\{c_{l,\alpha,\downarrow}(\mathbf{k})\},
\{c^{\dagger}_{l,\alpha,\uparrow}(-\mathbf{k})\},
\{c^{\dagger}_{l,\alpha,\downarrow}(-\mathbf{k})\}
\right]^{\mathrm T}
\end{equation*}
to obtain 
\begin{align}
H_{\mathrm{MF}}
&=
\frac{1}{2}
\sum_{\mathbf{k}}
\Psi^{\dagger}(\mathbf{k})
H_{\mathrm{BdG}}(\mathbf{k})
\Psi(\mathbf{k})
+
\frac{1}{2}
\sum_{\mathbf{k}}
\operatorname{Tr}\!\left[H_0(\mathbf{k})\right]
\nonumber\\
&\quad
+
\sum_{\mathbf{j},l}
\frac{|\Delta_{l,\mathbf{j}}|^2}{U}
\label{eq:HMF}
\end{align}
where $H_0(\mathbf{k})$ is the $PM \times PM$ single-particle Bloch Hamiltonian in momentum space and $H_{BdG}$ is the $4PM\times4PM$ BdG Hamiltonian which takes the form~\cite{deGennes1966,Schirmer2024}
\begin{equation*}
H_{\mathrm{BdG}}(\mathbf{k})
=
\begin{pmatrix}
H_{0}(\mathbf{k}) & 0 & 0 & -\Delta \\[2pt]
0 & H_{0}(\mathbf{k}) & \Delta & 0 \\[2pt]
0 & \Delta^{*}
& -H_{0}^{*}(-\mathbf{k}) & 0 \\[2pt]
-\Delta^{*} & 0 & 0
& -H_{0}^{*}(-\mathbf{k})
\end{pmatrix}
\end{equation*}
where $\Delta$ is the $PM\times PM$ on-site superconducting pairing matrix.

The self-consistent solution for the order parameter is obtained by the following iterative process: \begin{enumerate*}[label=(\roman*), itemjoin={{ }}, itemjoin*={{ }}]
    \item start with an initial guess $\Delta^{guess}_{l, \alpha}$, where $l\in\{1, 2,\ldots, P\}$ and $\alpha \in \{1,2,\ldots,M\}$ in the unit cell;
    \item construct the BdG Hamiltonian $H_{BdG}(\mathbf{k})$ using $\Delta^{guess}_{l, \alpha}$;
    \item compute the new order parameter $\Delta^{new}_{l, \alpha}$ by calculating the thermal average in Eq.~\eqref{eq:order_parameter}~\cite{Gorkov1958, Schirmer2024} and use this as the new guess;
    \item repeat this iteration until a self-consistent solution is obtained, defined by the convergence criterion 
    \end{enumerate*}
\begin{equation*}
    \max\limits_{l,\alpha}
    \left[
    \frac{
    \left|
    \Delta_{l,\alpha}^{(n+1)}
    -
    \Delta_{l,\alpha}^{(n)}
    \right|
    }{
    \left|
    \Delta_{l,\alpha}^{(n)}
    \right|
    }
    \right]
    \leq 10^{-5}
\end{equation*}
where $n$ is the iteration number.
\begin{figure*}[t!]
\centering


\begin{tikzpicture}
\node[inner sep=0] (a) {
    \includegraphics[
        width=0.47\textwidth,
        height=9cm,
        trim=1.5cm 5.0cm 2.0cm 4.8cm,
        clip
    ]{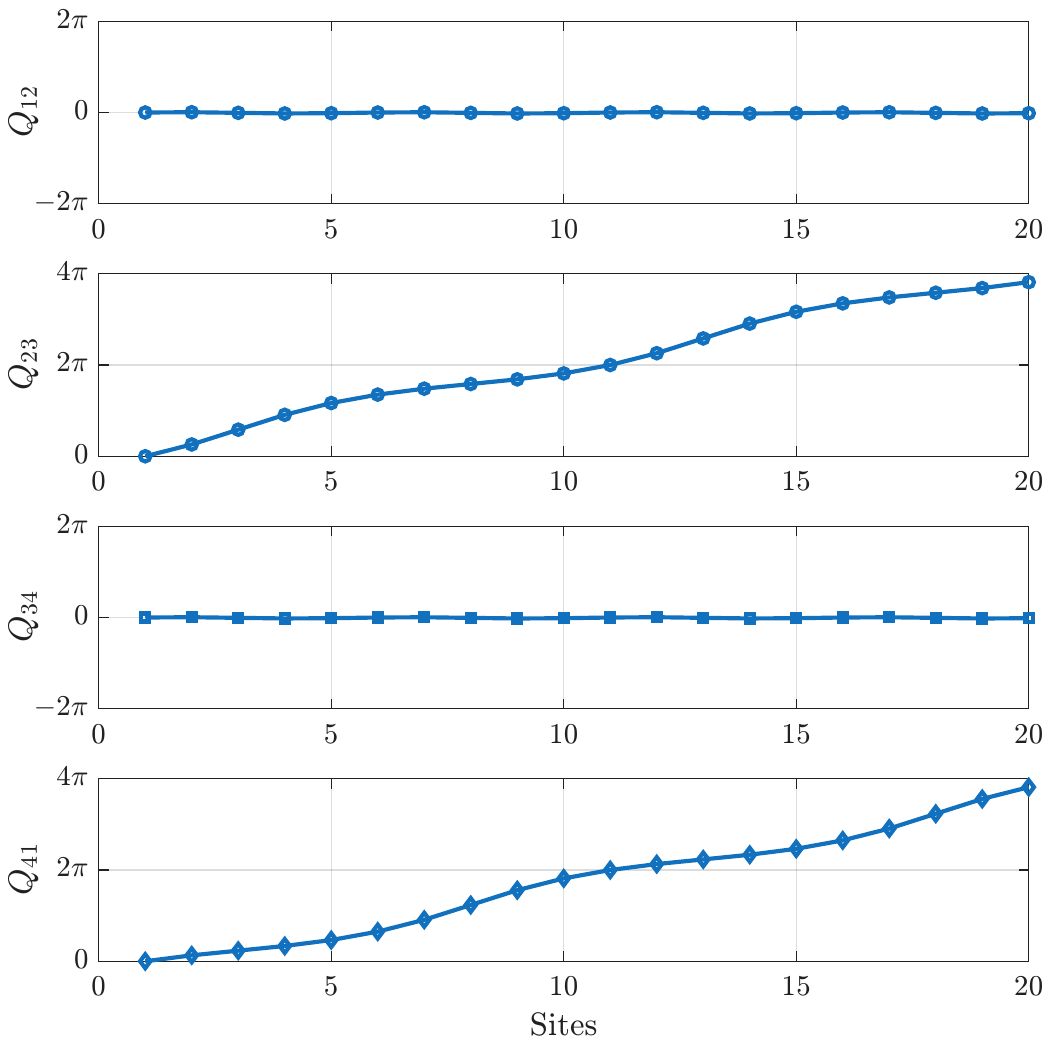}
};
\node[anchor=south,font=\large]
    at ([yshift=-0.05mm]a.north) {(a)};
\end{tikzpicture}
\hfill
\begin{tikzpicture}
\node[inner sep=0] (b) {
    \includegraphics[
        width=0.52\textwidth,
        height=8.8cm,
        trim=1.6cm 7.1cm 2.3cm 7.2cm,
        clip
    ]{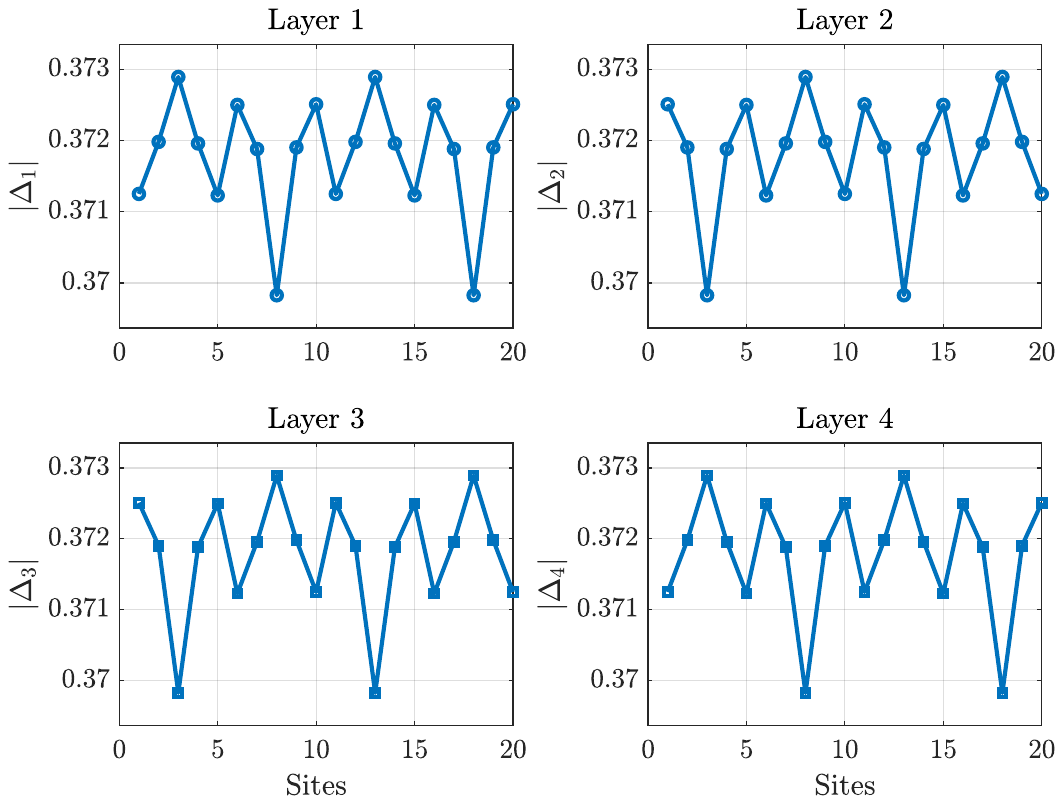}
};
\node[anchor=south,font=\large]
    at ([yshift=-0.001mm]a.north) {(b)};
\end{tikzpicture}

\par\vspace{0.15cm}


\begin{tikzpicture}
\node[inner sep=0] (c) {%
    \includegraphics[
        width=0.80\textwidth,
        trim=0.9cm 8.2cm 1.5cm 11cm,
        clip
    ]{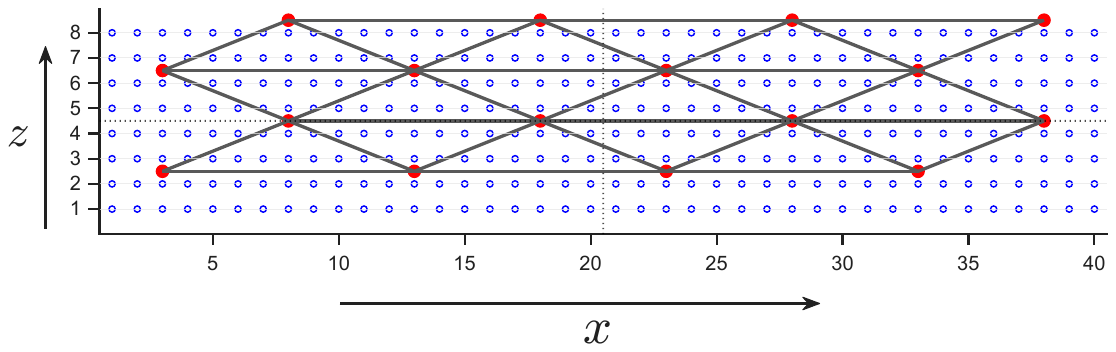}%
};

\node[anchor=north west,font=\large]
    at ([xshift=0.5mm,yshift=-0.5mm]c.north west) {(c)};
\end{tikzpicture}

\vspace{-1.2cm}

\caption{
(a) Gauge-invariant relative phases between adjacent layers of a
$20\times4$ MUC in the $xz$ plane. The $2\pi$ phase-windings identify two Josephson vortices between layers $2$ and $3$
and two across the periodic boundary between layers $4$ and $1$.
(b) Layer-resolved superconducting gap amplitudes $|\Delta_l|$ over
the same MUC. The interlayer vortices identified from the phase
windings in (a) are accompanied by local minima in the gap amplitudes
of the adjacent layers. For constructing the vortex configuration in
(c), we place each interlayer vortex at the $x$ position of the
corresponding gap-amplitude minimum and midway between the two
neighboring layers.
(c) Distorted triangular Josephson-vortex lattice obtained by
periodically repeating the MUC in the $x$ and $z$ directions.
The numerical parameters chosen for the calculation are $U=4$, $\mu=-3.5$, $t=0.25$,
$\Phi_p=0.025\Phi_0$, $N=80$, and $T=0$.
}
\label{fig:IL_vortex}

\end{figure*}
The initial guesses were typically of the form $\Delta^{\mathrm{guess}}_{l,\alpha}=\exp\left(2i\pi\alpha n_l/M\right)$, with $0 \leq n_l \leq \mathrm{int}(2\Phi/\Phi_0)+1$, where $\Phi$ is the total flux through the unit cell. Among the converged self-consistent states obtained from the
different initial configurations, we select the one with the lowest
free energy $F=\langle H_{MF}-TS\rangle$ as the thermodynamically stable solution at temperature $T$ with 
\begin{equation*}
\begin{aligned}
\left\langle H_{\mathrm{MF}}\right\rangle
&=
\frac{1}{2}
\sum_{\mathbf{k},\beta}
E_{\beta}(\mathbf{k})\,
n_{\mathrm{F}}\!\left(E_{\beta}(\mathbf{k})\right)
+
\frac{1}{2}
\sum_{\mathbf{k}}
\operatorname{Tr}\!\left[H_{0}(\mathbf{k})\right]
\\
&\qquad
+
\sum_{\mathbf{j},l}
\frac{\left|\Delta_{l,\mathbf{j}}\right|^{2}}{U}.
\end{aligned}
\end{equation*}
and
\begin{align*}
\langle S\rangle
&=
-k_{\mathrm{B}}
\sum_{\mathbf{k},\beta}
\Big[
n_{\mathrm{F}}\!\left(E_{\beta}(\mathbf{k})\right)
\ln n_{\mathrm{F}}\!\left(E_{\beta}(\mathbf{k})\right)
\nonumber\\
&\qquad
+
\left\{1-n_{\mathrm{F}}\!\left(E_{\beta}(\mathbf{k})\right)\right\}
\ln\!\left\{1-n_{\mathrm{F}}\!\left(E_{\beta}(\mathbf{k})\right)\right\}
\Big]
\end{align*}
where \(n_{\mathrm{F}}(E)=\left(e^{E/(k_{\mathrm{B}}T)}+1\right)^{-1}\) is the Fermi function and $\beta$ labels the BdG bands. In practice, different initial guesses $\Delta_{l,\alpha}^{\mathrm{guess}}$ often converge to the same self-consistent solution corresponding to the free energy minima.

Writing the superconducting order parameter as
$\Delta_{l,j}=|\Delta_{l,j}|e^{i\phi_{l,j}}$, we define the
gauge-invariant relative phase between neighboring layers as (closely following Ref.~\cite{Nag2025})
\begin{equation}
Q_{l,l+1}(j)
=
\phi_{l,j}-\phi_{l+1,j}
+2A^l_{z,j}
\end{equation}
Here, the factor of two accounts for the Peierls phase accumulated by the two electrons forming a Cooper pair. A $2\pi$ winding of $Q_{l,l+1}(j)$ along the
$x$ direction identifies an interlayer Josephson vortex. For the infinite-layer geometry, the same definition is applied
across the MUC boundary between layers $P$ and $1$ of the
neighboring unit cell.

Finite systems with APBCs can
support a net current that vanishes in the
thermodynamic limit~\cite{Watanabe2019}. To reduce this finite-size
effect, we introduce a uniform boundary twist $q$ along
$\hat{\mathbf{x}}$, such that the boundary condition now becomes
\begin{equation}
c_{l,\mathbf{j}+N\hat{\mathbf{x}},\sigma}
=
-e^{iqN}c_{l,\mathbf{j},\sigma}
\end{equation}
The derivative of the free energy with respect to the twist is
related to the total current along $\hat{\mathbf{x}}$ by
\cite{Watanabe2019}
\begin{equation}
\frac{\partial F}{\partial q}
=
\sum_{l,\mathbf{j}}
\left\langle
\hat{J}^{\,l}_{x,\mathbf{j}}
\right\rangle 
\label{eq:twist}
\end{equation}
We therefore minimize the free energy with respect to $q$, which
ensures that the net current along $\hat{\mathbf{x}}$ vanishes and
yields a self-consistent solution that more closely represents the
zero-current thermodynamic limit, consistent with Bloch's theorem
for equilibrium lattice systems~\cite{Watanabe2019}. The explicit
definitions of the local supercurrents and related numerical details are presented
in Appendix~\ref{app:currents}.\\

{\section {Vortex lattice in the infinite-layer geometry}\label{sec:infinite}}
We first examine the spatial structure of a representative
self-consistent ground state in the infinite-layer geometry. Fig.~\ref{fig:IL_vortex}(a)
shows the gauge-invariant relative phases between neighboring layers
within a $20\times4$ MUC in the $xz$ plane. For the state shown here, two vortices occur between
layers $2$ and $3$, while two additional vortices occur between layers $4$ and $1$ of the neighboring MUC;
no vortices are present in the other interlayer junctions. The corresponding
layer-resolved gap amplitudes are shown in Fig.~\ref{fig:IL_vortex}(b). Each phase
winding is accompanied by a local suppression of the superconducting
gap in the adjacent layers, allowing the vortex position along
$\hat{\mathbf{x}}$ to be associated with the corresponding local gap
minimum up to a spatial resolution of one lattice spacing
along both $\hat{\mathbf{x}}$ and $\hat{\mathbf{z}}$. Repeating the MUC periodically along both $\hat{\mathbf{x}}$
and $\hat{\mathbf{z}}$ then produces the vortex configuration shown in
Fig.~\ref{fig:IL_vortex}(c), where the Josephson vortices form a distorted
triangular lattice. The self-consistent solution has been verified using a larger $40\times4$ MUC in the $xz$ plane. Triangular Josephson-vortex lattices are expected for layered superconductors in a magnetic field parallel to the superconducting planes~\cite{Bulaevskii1991,Balents1996,Yu2007}; thus, our self-consistent BdG procedure reproduces the expected infinite-layer vortex structure while additionally resolving the layer-dependent gap amplitude and phase profile microscopically.

A triangular lattice is expected in the continuum limit for an isotropic system. The deviation from a triangular lattice primarily reflects
the strong mass anisotropy between the in-plane and interlayer directions.
Within anisotropic GL theory, the vortex lattice becomes triangular after rescaling the stacking direction by the effective-mass anisotropy ($\gamma$) $z \rightarrow \gamma z$, while leaving the in-plane coordinates unchanged~\cite{Campbell1988, Petzinger1990}. Using the bare
electron masses obtained from the quadratic band-bottom dispersion of
our tight-binding model, with the lattice
spacings set to unity, gives
$\gamma=\sqrt{m_z/m_x}=\sqrt{1/t_2}$. For the parameters of
Fig.~\ref{fig:IL_vortex}, this rescaling ($\gamma=2$) expands the lattice in the direction required
to approach an equilateral triangle, although the resulting lattice
remains noticeably distorted. The relevant anisotropy factor in the GL description, however, is associated with the stiffness of the superconducting order parameter, equivalently with an effective Cooper-pair mass, rather than directly with the bare electronic band masses~\cite{Koshelev2013, Iazzi2012}. In the strong-coupling limit of the attractive Hubbard model,
the effective pair-hopping amplitude along direction $i$ is
$J_i \simeq 2t_i^2/U$~\cite{Micnas1990}.
Since the effective pair mass scales as
$m_i^{\rm pair}\propto J_i^{-1}$, this gives
$\gamma_{\rm pair}
=\sqrt{m_z^{\rm pair}/m_x^{\rm pair}}
\simeq t_1/t_2=4$. In the moderately
strong-pairing regime considered here, estimates based on the effective
pair hopping give a slightly reduced anisotropy factor
$\gamma\simeq3.7$-$4$ (see Appendix~\ref{app:anisotropy} for more details), for which the
rescaled lattice is nearly equilateral. The remaining deviation arises from the four-layer MUC, which constrains the allowed
vortex separation along the stacking direction, as well as from the
discrete spatial resolution of the vortex positions in our lattice calculation.\\

{\section {Phase diagrams}\label{sec:pdiag}}
\begin{figure*}[t]
    \centering

    \includegraphics[width=0.332\textwidth]{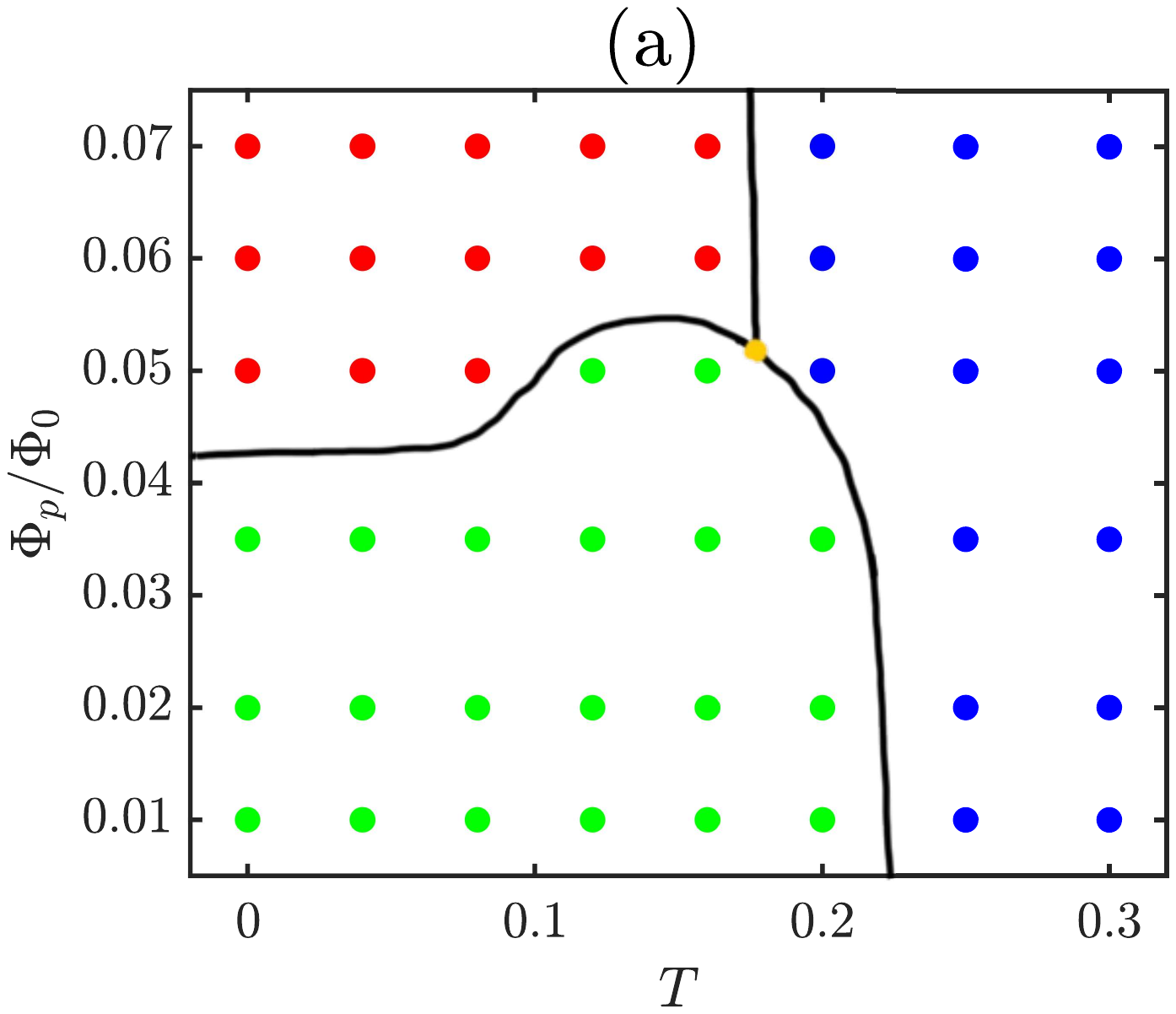}%
    \includegraphics[width=0.332\textwidth]{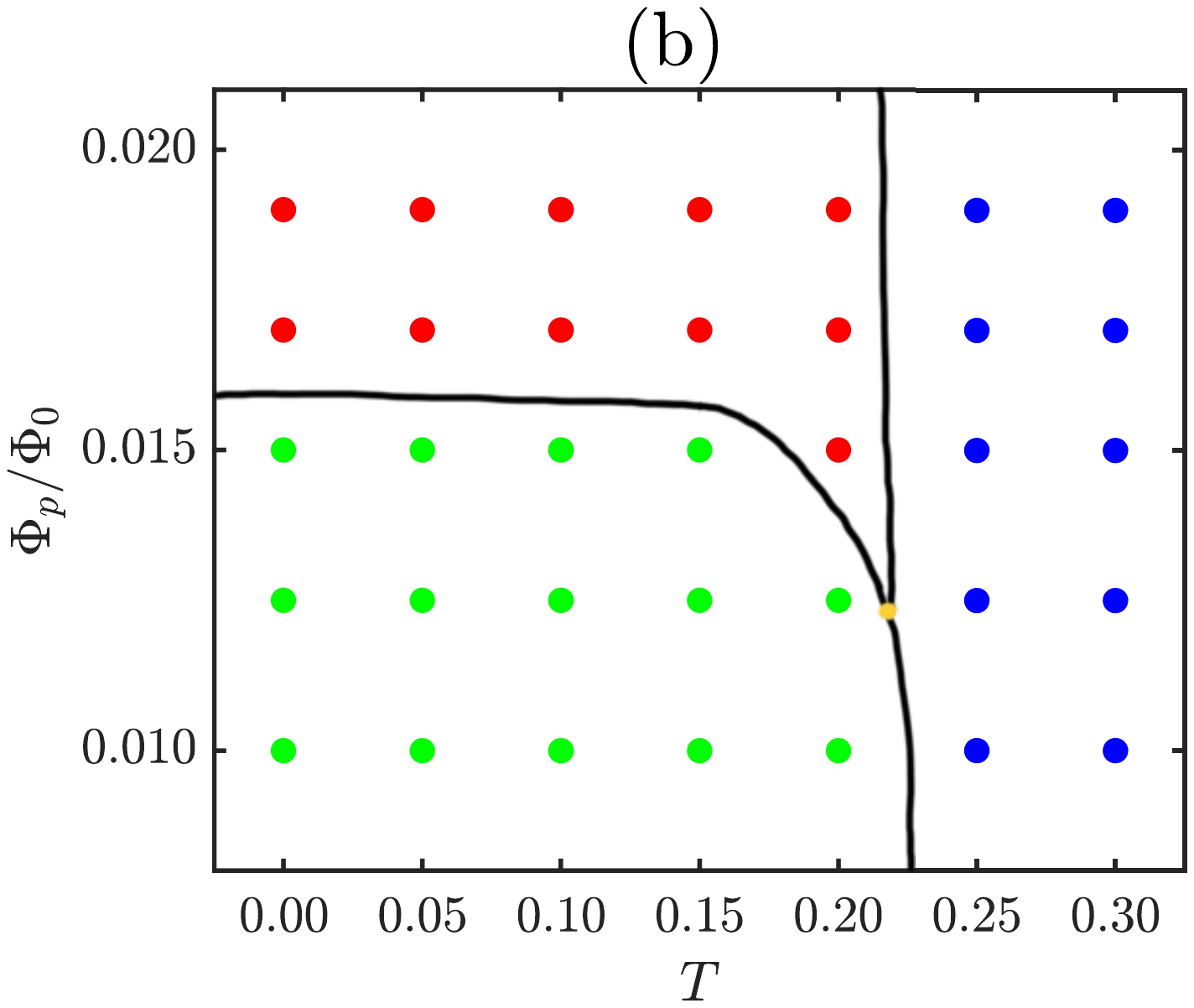}%
    \includegraphics[width=0.332\textwidth]{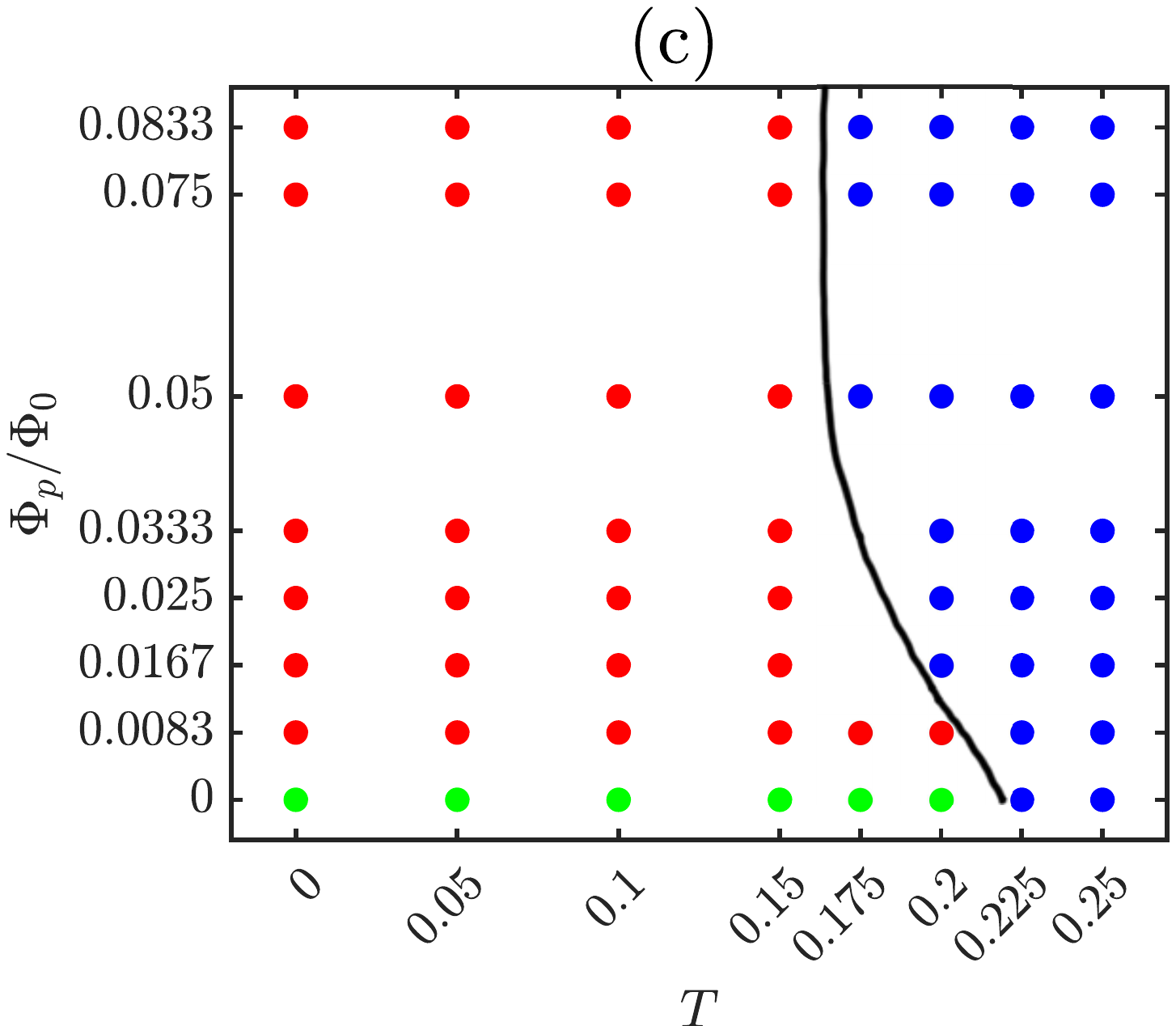}

    \vspace{-0.15cm}

    \makebox[\textwidth][c]{%
        \includegraphics[width=0.46\textwidth]{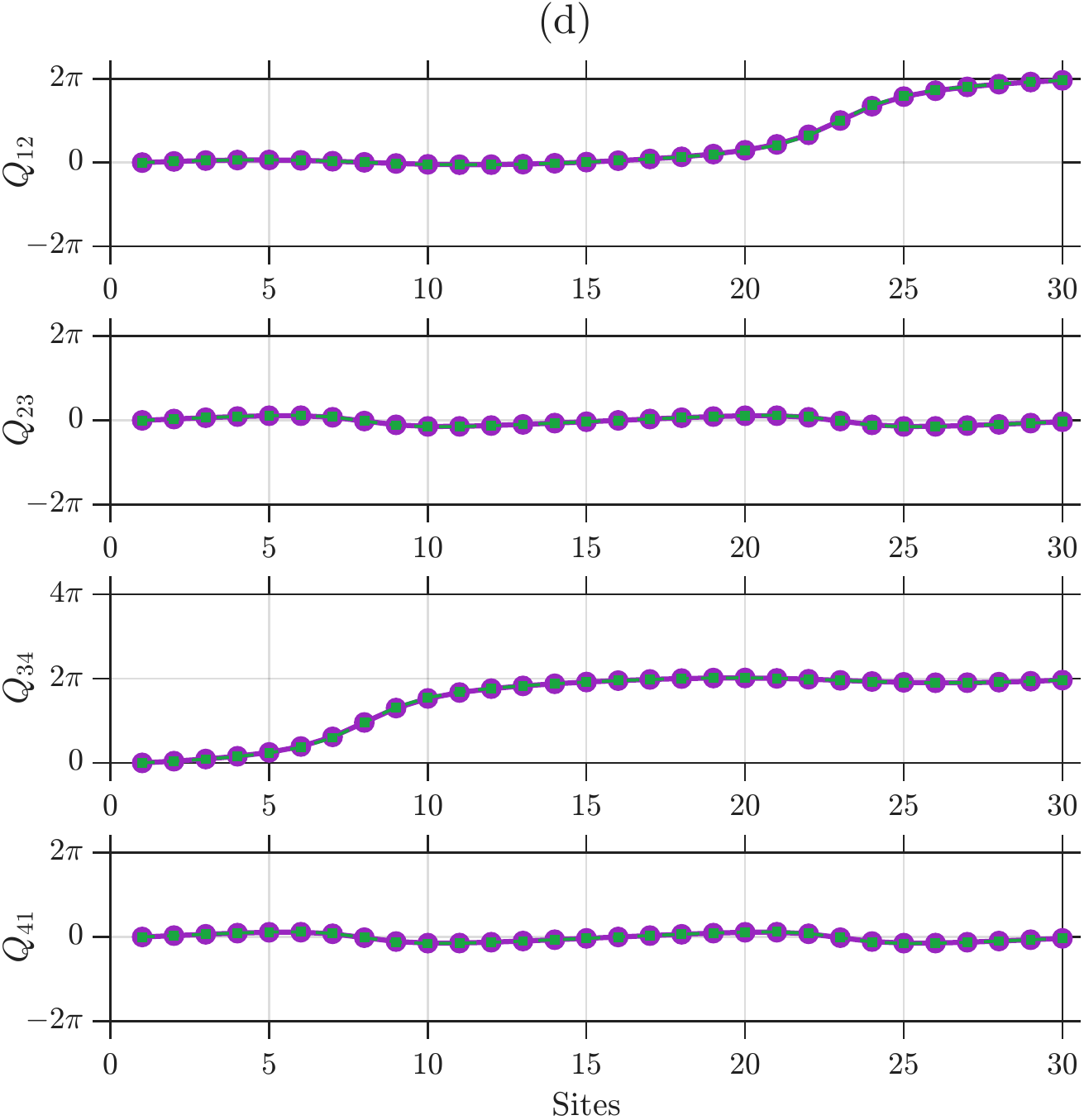}%
        \hspace{0.015\textwidth}%
        \raisebox{8mm}{%
    \includegraphics[width=0.50\textwidth]
        {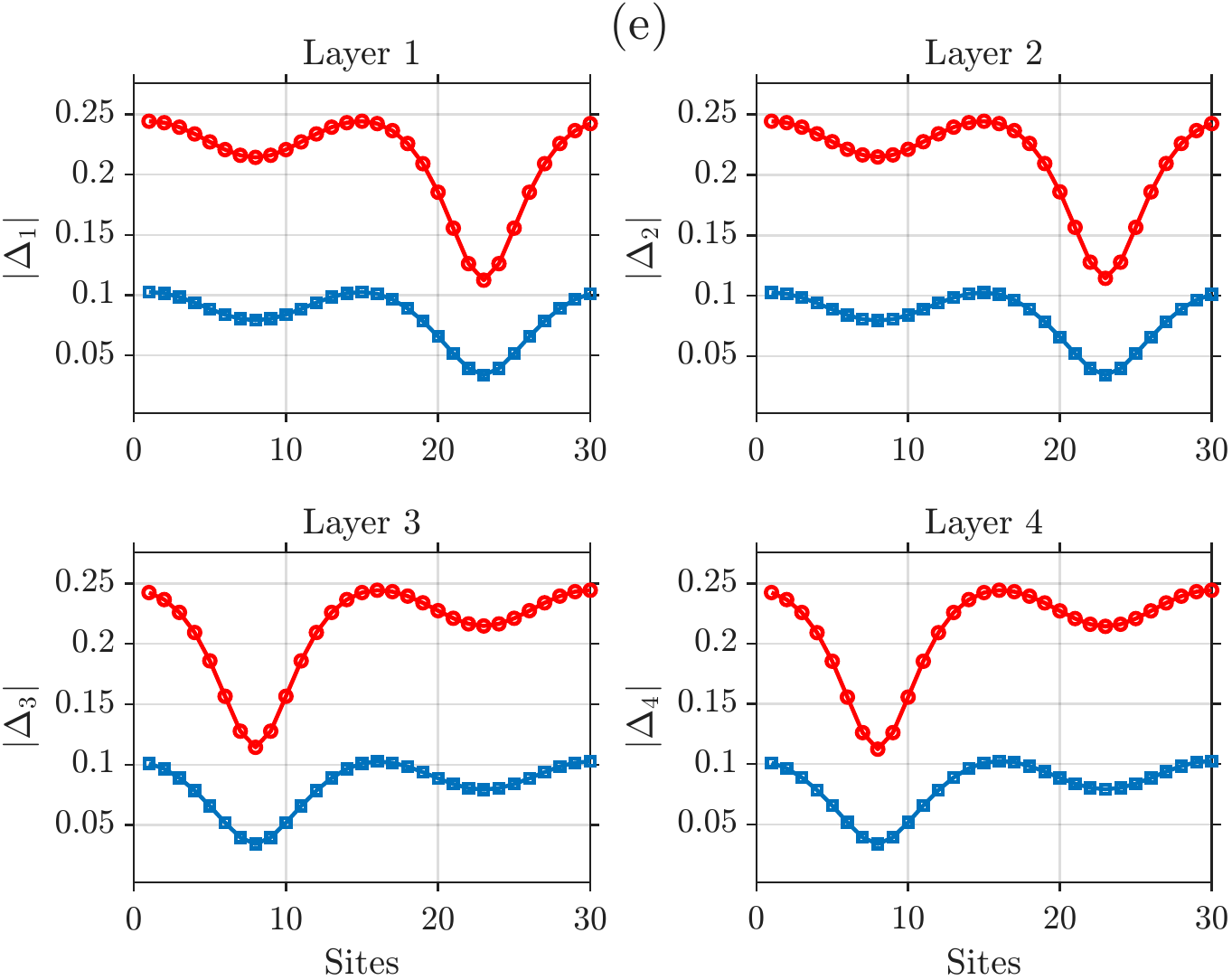}%
}

    }

    \vspace{-0.15cm}

    \caption{
Temperature--flux phase diagrams for the
(a) bilayer, (b) four-layer, and (c) infinite-layer systems.
Green, red, and blue symbols denote the BCS, orbital-FFLO,
and normal phases, respectively.
The numerical parameters for (a) and (b) are
$U=4$, $\mu=-3.5$, $t=0.25$, and $N=40$, while for
(c) we use $N=60$ with the remaining parameters unchanged. The tricritical point shifts to lower flux and higher temperature from (a) to (b), while the BCS region is progressively suppressed and disappears in the infinite-layer limit for $\Phi_p>0$, where the low-field orbital-FFLO–-normal phase boundary upturn persists. (d) Gauge-invariant relative phases and
(e) layer-resolved superconducting gap amplitudes in the
infinite-layer system for $\Phi_p=0.0083\Phi_0$ at $T=0.175$ (red curve) and $T=0.200$ (blue curve), two points close to $T_{c0}$
below the phase boundary upturn.
The relative-phase profiles at these two temperatures coincide in (d), indicating that the phase-winding structure is unchanged, while the gap amplitudes are substantially reduced as the phase boundary is approached. The calculation is shown for a $30\times4$ MUC in the $xz$ plane.
}
    \label{fig:phase_diagrams}
\end{figure*}

We next examine how the superconducting phase diagram evolves upon going
from the bilayer to the infinite-layer limit. Fig.~\ref{fig:phase_diagrams} shows the
temperature-flux phase diagrams for the (a) bilayer, (b) four-layer,
and (c) infinite-layer systems. A clear
trend is the progressive suppression of the uniform BCS phase with
increasing layer number. While the bilayer system supports a substantial BCS
region at low fields, this region is strongly reduced in the four-layer
system and is essentially absent at nonzero magnetic flux in the infinite-layer calculation over the parameter regime considered here. The accompanying shift of the tricritical point toward lower fields and  higher temperatures is qualitatively consistent with finite-layer GL theory~\cite{Yan2026,Yuan2023} and the thickness-dependent experimental trend reported in Ref.~\cite{Cao2026}. To understand the suppression of the BCS phase with increasing layer number from a phenomenological viewpoint, we follow the GL argument of~\cite{Yan2026} and make the layer-number dependence explicit by considering the orbital energy cost of imposing a common phase gradient on all layers. As derived in
Appendix~\ref{app:bcs_suppression}, for an inversion-symmetric
stack of $P$ equally spaced layers with interlayer spacing $d$,
the orbital energy cost per layer of maintaining the uniform BCS
state scales as
\begin{equation}
\frac{\Delta E_{\mathrm{orb}}}{P}
\propto
\left(\frac{2eBd}{\hbar}\right)^2
\frac{P^2-1}{12}
\label{eq:bcs_orbital_cost}
\end{equation}
The above equation shows that $\Delta E_{\mathrm{orb}}/P$ grows rapidly
with increasing layer number and diverges in the infinite-layer
limit. Therefore, allowing layer-dependent Cooper-pair momenta becomes
increasingly favorable, naturally explaining the progressive
suppression of the BCS region in favor of the orbital-FFLO
state.

Within the GL framework of Refs.~\cite{Yan2026, Schneider1993}, the enhancement of the upper-critical field $H_{c2}$ on reducing the temperature from the zero-field critical temperature $T_{c0}$ has been associated with a dimensional crossover controlled by the out-of-plane coherence length $\xi_c(T)$. Close to $T_{c0}$, $\xi_c(T)\propto(1-T/T_{c0})^{-1/2}$ exceeds the interlayer spacing $D$, so the superconducting order parameter remains coherent across several neighboring layers. As the temperature is lowered, $\xi_c(T)$ decreases, and the crossover $\xi_c(T^*)\sim D$ marks the onset of a more layer-confined (2D) regime. This crossover produces the characteristic low-field upturn of the orbital-FFLO--normal phase boundary. Our microscopic BdG results in Fig.~\ref{fig:phase_diagrams}(d) and Fig.~\ref{fig:phase_diagrams}(e) provide a layer-resolved view of the superconducting state close to $T_{c0}$ and below the upturn. At $\Phi_p=0.0083\Phi_0$, the gauge-invariant relative-phase
profiles at $T=0.175$ and $T=0.200$ essentially coincide,
indicating that the phase-winding structure is unchanged between
the two temperatures. The corresponding gap-amplitude profiles
also show minima at the same positions, confirming that the
vortex configuration remains unchanged. The main
evolution instead occurs in the overall gap amplitude: as the orbital-FFLO--normal phase boundary is approached, $|\Delta_l|$ is strongly reduced, while the vortex-associated suppression continues to extend across neighboring layers indicating a 3D superconducting regime. This behavior is consistent with the GL crossover picture, although our BdG calculation does not directly extract $\xi_c(T)$. A direct microscopic characterization of the crossover at higher fields is complicated by the strong overlap of gap suppressions associated with neighboring vortices, making it difficult to isolate the contribution of an individual vortex in a four-layer MUC.\\

{\section{Vortex rearrangement in the four-layer system}\label{sec:rearrangement}}

\begin{figure}[t]
    \centering
    \includegraphics[width=\columnwidth]{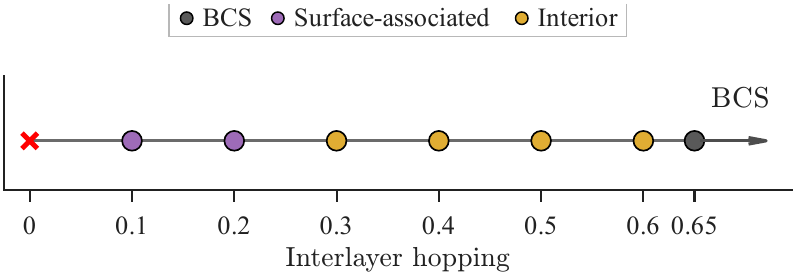}

    \caption{
    Evolution of the ground state with interlayer hopping $t$ in four-layer systems for $\Phi_p=0.030\Phi_0$, $N=20$, $U=4$, $\mu=-3$, and $T=0$. With increasing $t$, the vortex configuration changes from surface-associated to interior, followed by a transition to
    the uniform BCS state at larger $t$. The $t=0$ point is left unclassified because the layers are completely decoupled in this limit. “Surface-associated” and “interior” denote configurations with vortices localized near the outer and inner layers, respectively.
    }
    \label{fig:4layer_tscan}
\end{figure}
\begin{figure*}[t]
    \centering

    \begin{minipage}[t]{0.48\textwidth}
        \centering
        \begin{overpic}[width=\linewidth]
            {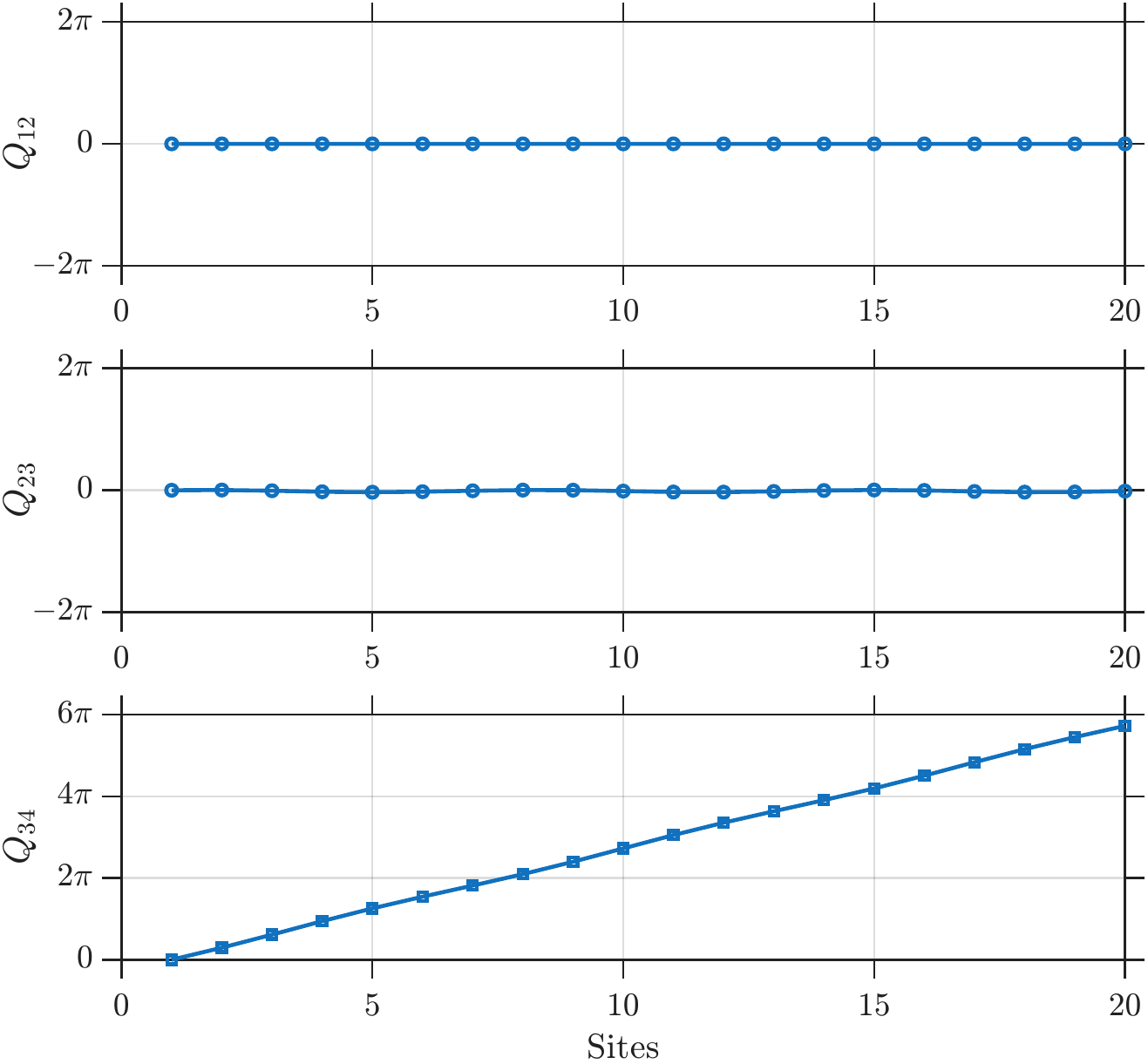}
            \put(2,97){\large(a)}
        \end{overpic}
    \end{minipage}
    \hfill
    \begin{minipage}[t]{0.48\textwidth}
        \centering
        \begin{overpic}[width=\linewidth]
            {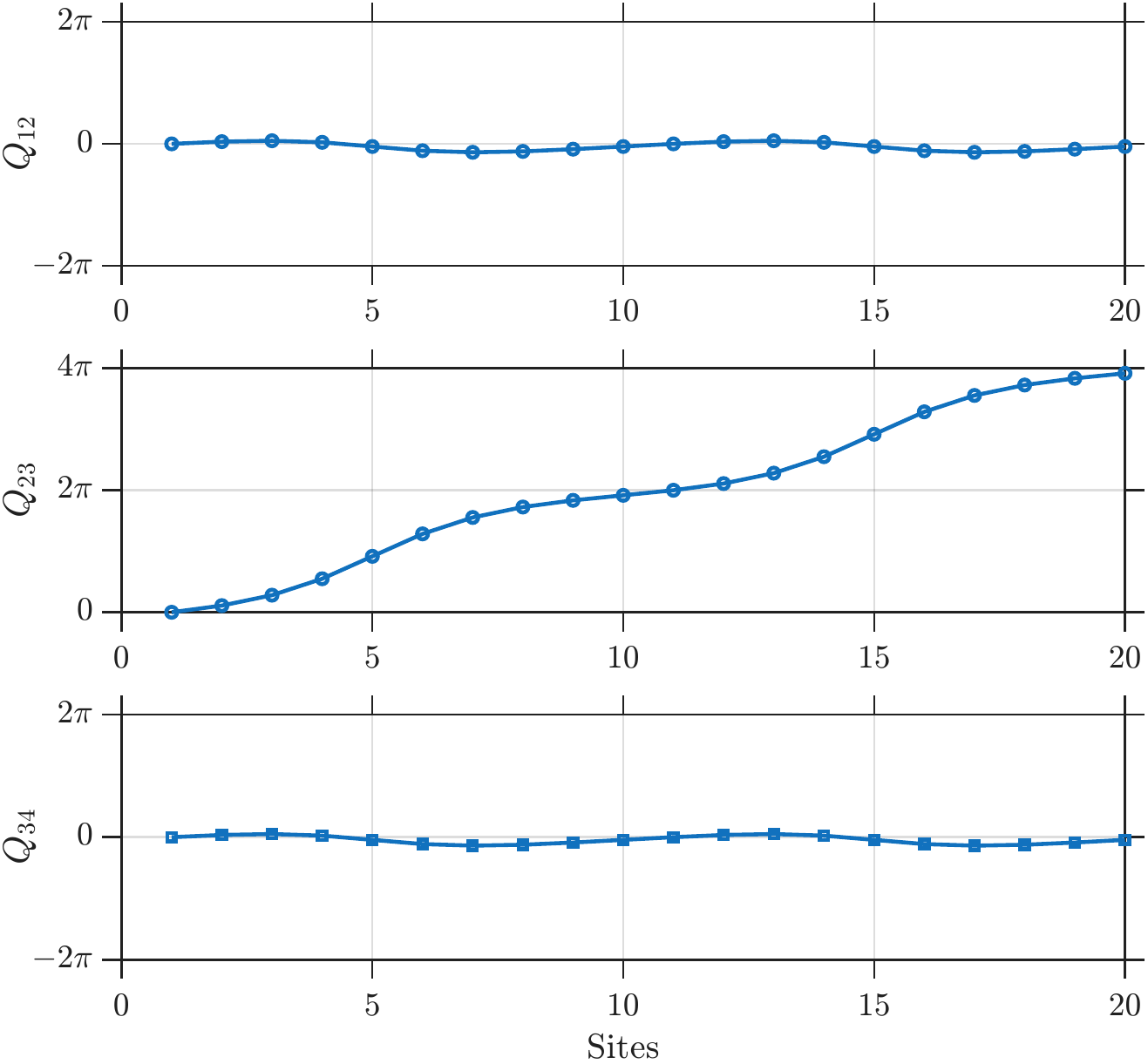}
            \put(2,97){\large(b)}
        \end{overpic}
    \end{minipage}

    \vspace{0.6em}

    \begin{minipage}[t]{0.48\textwidth}
        \centering
        \begin{overpic}[width=\linewidth]
            {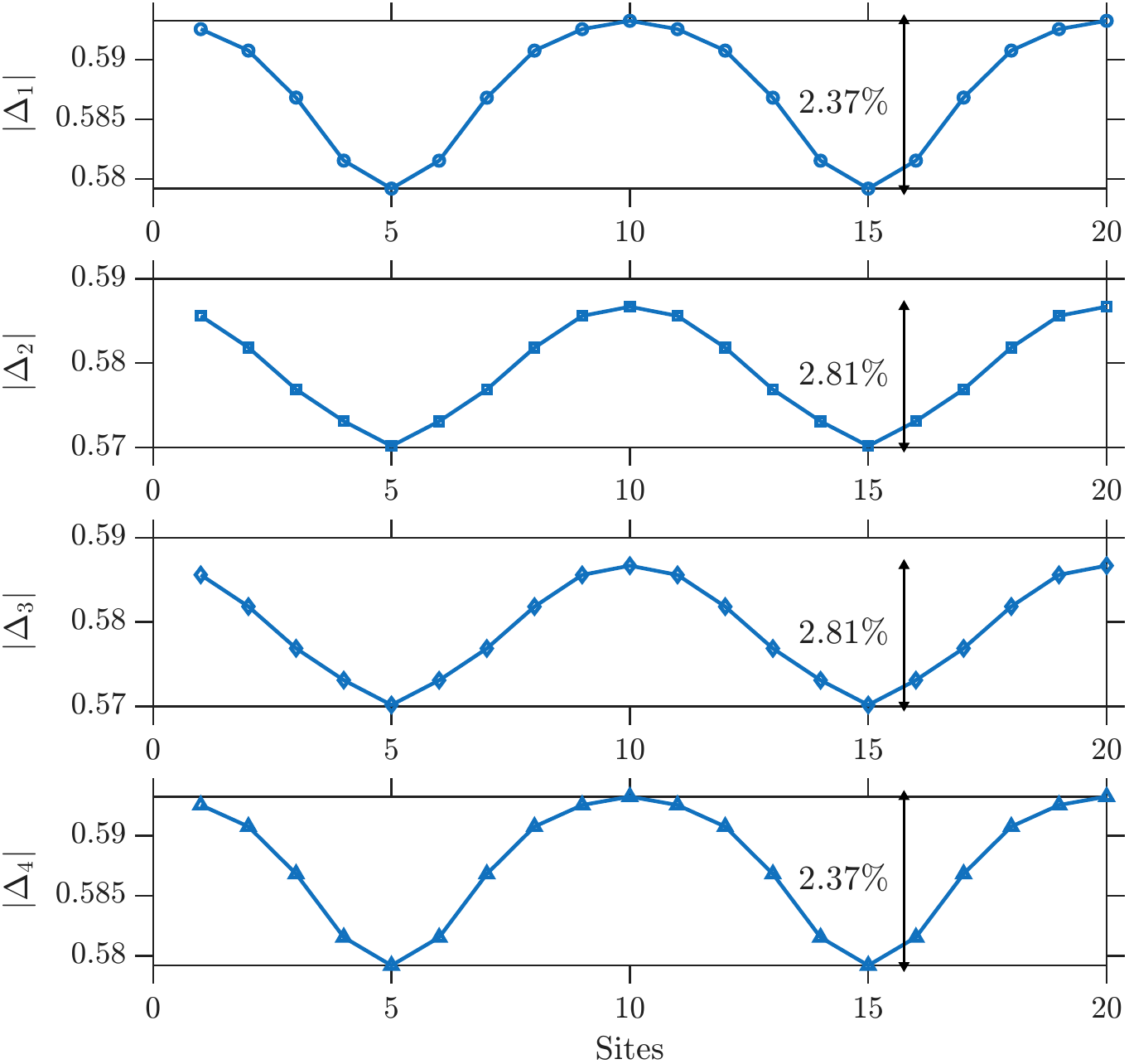}
            \put(-10,83){\large(c)}
        \end{overpic}
    \end{minipage}

    \caption{
Rearrangement of Josephson vortices with increasing interlayer
hopping in the four-layer system at
$\Phi_p=0.030\Phi_0$, $N=20$, $U=4$, $\mu=-3$, and $T=0$.
We show two representative points from the interlayer-hopping
phase diagram in Fig.~\ref{fig:4layer_tscan}.
(a) Gauge-invariant interlayer relative phases
$Q_{l,l+1}(j)$ at $t=0.20$. The $6\pi$ winding of
$Q_{34}$ corresponds to three Josephson vortices in the
$3$--$4$ interlayer gap, giving an edge-localized
vortex configuration.
(b) Relative phases at $t=0.40$. The $4\pi$ winding of
$Q_{23}$ corresponds to two Josephson vortices in the
interior $2$--$3$ gap, illustrating the rearrangement
of vortices toward the interior with increasing interlayer hopping.
(c) Layer-resolved superconducting order-parameter magnitudes
$|\Delta_l(j)|$ for the $t=0.40$ configuration. Although the
vortices reside in the interior $2$--$3$ gap, the outer
layers exhibit a suppression of the gap amplitude comparable to
that of the interior layers.
}
    \label{fig:4layer_rearr}
\end{figure*}
Fig.~\ref{fig:4layer_tscan} shows the interlayer hopping $t$ dependence of the vortex structure for a representative four-layer system, illustrating how the Josephson-vortex arrangement evolves from surface-associated to interior configurations with increasing $t$. To examine
the microscopic structure of these states, in
Fig.~\ref{fig:4layer_rearr} we show two representative points from
Fig.~\ref{fig:4layer_tscan}, at $t=0.20$ and $t=0.40$, for
$\Phi_p=0.030\Phi_0$, $N=20$, and $T=0$. The corresponding
energy-minimizing boundary twists are $q=-0.03684$ and $q=0$,
respectively. Figure~\ref{fig:4layer_rearr}(a) shows the
gauge-invariant relative phases $Q_{12}$, $Q_{23}$, and $Q_{34}$
for the surface-associated state at $t=0.20$. The $6\pi$ winding
of $Q_{34}$ corresponds to three Josephson vortices in the outer
$3$--$4$ interlayer gap. The inversion-related configuration, with the vortices located in the
opposite outer interlayer junction, is degenerate with the state shown
in Fig.~\ref{fig:4layer_rearr}(a). Figure~\ref{fig:4layer_rearr}(b)
shows the interior configuration at $t=0.40$, for which the
$4\pi$ winding of $Q_{23}$ corresponds to two Josephson vortices
in the interior $2$--$3$ gap.
Figure~\ref{fig:4layer_rearr}(c) shows the layer-resolved gap
amplitudes $|\Delta_l|$ for the $t=0.40$ interior-vortex
configuration. Despite the vortices residing in the interior
$2$--$3$ junction, the gap modulation remains appreciable at the
surface: the suppression is approximately $3\%$ in the two
inner layers and $2.4\%$ in the two surface layers. Thus, the suppression of gap in the surface layers is expected to be comparable to that in the interior,
showing that the deformation associated with an interior
Josephson vortex extends across the four-layer stack. This
surface imprint provides a low temperature STM-accessible signature of
Josephson vortices residing below the surface. In Appendix~\ref{app:6L}, we present selected six-layer calculations demonstrating that vortices located deeper within a thicker finite stack still produce a finite, albeit much weaker, modulation of the surface-layer gap.

A qualitative physical interpretation of this rearrangement can be given
in terms of surface effects in the finite-layer geometry. At smaller $t$,
a surface-associated vortex can be less costly because a boundary layer is
coupled to only one neighboring layer, giving it greater freedom to
accommodate the field-induced phase deformation without requiring the
distortion to propagate throughout the entire stack. As $t$ increases,
the stronger interlayer coupling favors a more collective deformation
across the stack, and in our four-layer calculations the lowest-energy
vortex configuration shifts toward the interior. In this configuration,
the phase and amplitude distortions can be distributed more symmetrically
between the two halves of the system, as evident from Fig.~\ref{fig:4layer_rearr}(c).

We performed extensive robustness checks on the finite-layer
solutions. In addition to minimizing the energy with respect to the boundary twist
and verifying the corresponding equilibrium-current condition
(Appendix~\ref{app:currents}), we examined the interlayer-hopping-driven
vortex rearrangement over additional magnetic fields and system sizes.
At a slightly lower flux than the main-text value, we carried out the
calculation for both $N=20$ and the larger system size $N=30$, providing
a direct check of the persistence of the rearrangement with increasing
system size. We also examined a larger magnetic flux for $N=20$. In all of these four-layer
calculations, increasing the interlayer hopping produces the same
qualitative reorganization of the vortex configuration
toward the middle interlayer junction; these results are presented in
Appendix~\ref{app:add_4L}. We also verified that the identified
configurations are insensitive to the choice of boundary conditions,
as discussed in Appendix~\ref{app:bc}. To analyze the local stability
of the solutions, we applied small random perturbations to the converged order
parameter and verified that the self-consistent iterations reconverge
to the same solution. We further evaluated the energy landscape under 2D
complex perturbations of representative four-layer solutions and found
a unique minimum at the unperturbed configuration
(Appendix~\ref{app:stability}). These numerical tests and consistency checks give us confidence that the configurations reported here represent the ground states for the parameter regimes considered.\\

{\section{Discussion and conclusions}\label{sec:discussion}}

A central result of our work is the progressive suppression of the uniform BCS region with increasing layer number. The BCS regime is substantial in the bilayer geometry, strongly reduced in the four-layer system, and essentially absent at nonzero flux in the infinite-layer calculation over the parameter range considered. This trend is consistent with the orbital-energy argument presented in Appendix~\ref{app:bcs_suppression} and with recent tunneling-spectroscopy measurements on multilayer $\mathrm{2H}\text{-}\mathrm{NbSe}_2$, which found that the transition field into the orbital-FFLO regime decreases strongly with increasing sample thickness~\cite{Cao2026}. In the infinite-layer system, we additionally find a low-field upturn of the orbital-FFLO--normal phase boundary. While finite-layer GL theory captures the dimensional-crossover mechanism underlying the phase boundary upturn, it does not provide the corresponding reduced crossover temperature. Our microscopic calculation allows this scale to be extracted directly, allowing a comparison with experiment. In our infinite-layer calculation, the upturn sets in at \(T^\ast=0.200\pm0.0125\), while the zero-field critical temperature is bracketed between \(T=0.200\) and \(0.225\), giving \(T_{c0}=0.2125\pm0.0125\). This corresponds to a reduced temperature \(T^\ast/T_{c0}\simeq0.94\pm0.08\), where the uncertainty reflects the finite temperature resolution of our phase diagram. In the experiments of Ref.~\cite{Wan2023}, the corresponding \(H_{c2}\) upturn in the thicker 34- and 39-nm $\mathrm{2H}\text{-}\mathrm{NbSe}_2$ samples occurs at \(T^\ast/T_{c0}\sim0.85\!-\!0.90\), as estimated from Fig. 1(i) of the article. These are also the samples for which the observed upturn was found to be consistent with a dimensional crossover picture. Thus, the reduced temperature scale of the upturn in our infinite-layer calculation is quantitatively in agreement (within the numerical temperature resolution) with that observed experimentally in the thick-sample regime. Hence, our microscopic BdG calculations provide a bridge between phenomenological GL descriptions of orbital-FFLO superconductivity~\cite{Liu2017,Yuan2023,Yan2026} and the microscopic bilayer picture developed in Ref.~\cite{Nag2025}, while also offering a framework for interpreting recent experimental trends in multilayer $\mathrm{2H}\text{-}\mathrm{NbSe}_2$~\cite{Wan2023,Cao2026}. In particular, the phase diagrams and layer-resolved vortex structures obtained here help connect the theoretical evolution with layer number to the behavior observed experimentally.

The layer-resolved BdG calculation further reveals an evolution of the Josephson-vortex configuration with interlayer hopping. Experimentally, the interlayer coupling in NbSe$_2$ can be tuned
through ionic intercalation, which produces a layer-decoupled,
effectively 2D electronic structure~\cite{Zhang2022Intercalated},
or by applying pressure to modify the interlayer spacing. For the four-layer systems examined here, the lowest-energy states evolve from surface-associated (edge-localized or staggered) vortex structures to vortex configurations concentrated in the interior as the hopping is increased. The same qualitative tendency toward an interior vortex distribution is found at additional fluxes and a larger system size in Appendix~\ref{app:add_4L}. We find that the interior vortices also produce a finite variation of the superconducting order parameter at the surface. The resulting surface modulation may therefore provide an STM-accessible signature of Josephson vortices located in the interior of the stack. Indeed, STM techniques have previously demonstrated that subsurface vortices in bulk $\mathrm{2H}\text{-}\mathrm{NbSe}_2$ produce detectable signatures in the surface LDOS~\cite{Fridman2011}. More recently, this connection was resolved at the level of vortex-bound Caroli--de Gennes--Matricon quasiparticle states, with the surface-projected spectrum shown to evolve systematically with the depth of the subsurface vortex~\cite{Xi2025HiddenVortices}. While this provides useful experimental precedent, the corresponding signatures in finite-layer systems remain to be established. A quantitative prediction for the tunneling signal in our four-layer systems would require calculation of the surface LDOS and is left for future study. Selected six-layer calculations in Appendix~\ref{app:6L} show that vortices located deeper within a thicker stack likewise produce a weaker surface-layer modulation, providing a qualitative extension of the surface-imprint result to a thicker finite stack. \\

{\it Acknowledgments:} 
UN was supported in part by the U. S. Department of Energy, Office of Basic Energy Sciences, under Grant no. DE-SC0005042. 
We acknowledge Advanced CyberInfrastructure computational resources provided by The Institute for CyberScience at The Pennsylvania State University. SM and UN thank the Lodha Theoretical Physics Institute, Mumbai, for their hospitality, where part of this work was done.\\
\begin{figure*}[t]
    \centering

    \begin{minipage}[t]{0.495\textwidth}
        \centering
        {(a)}\\[-0.3mm]
        \includegraphics[
            width=\linewidth,
            trim=30 120 35 145,
            clip
        ]{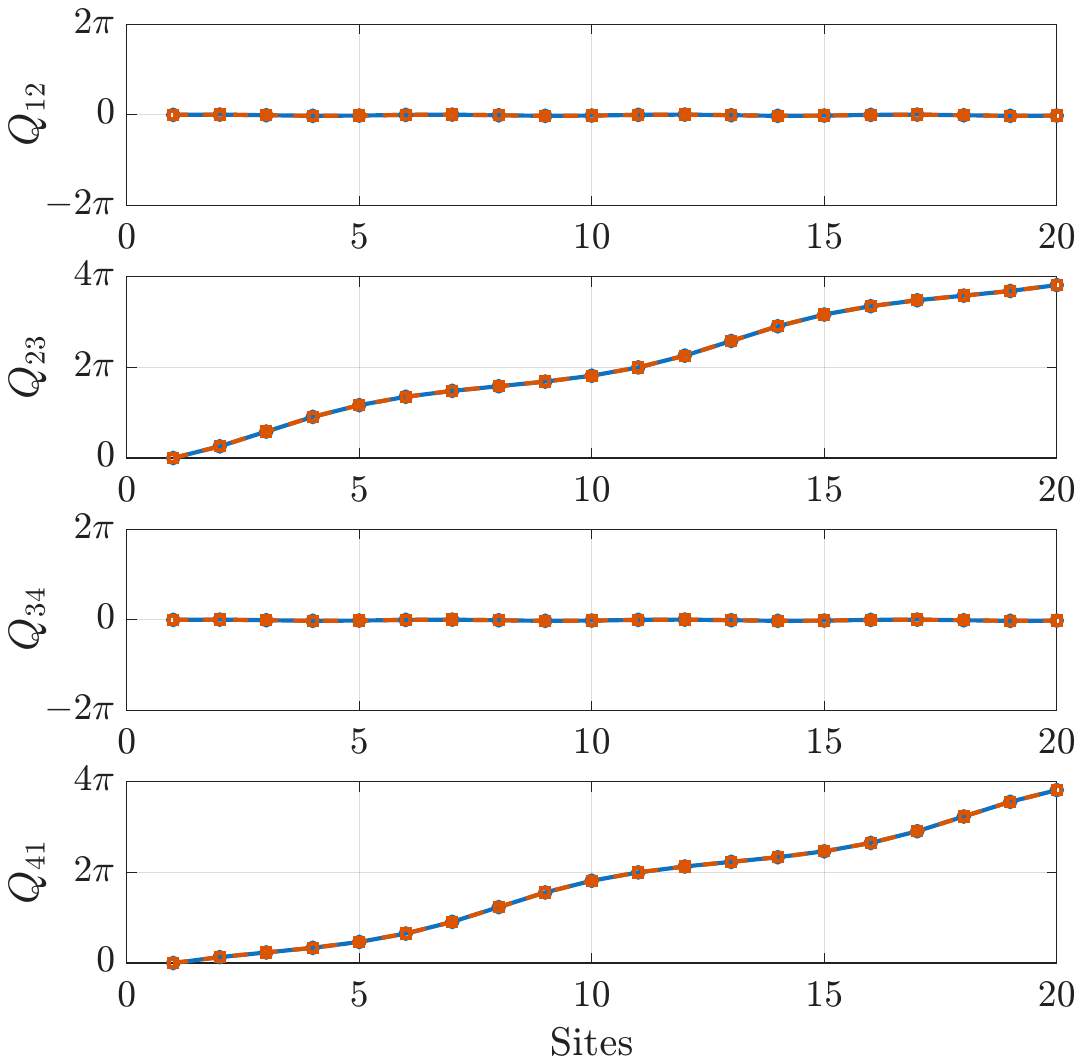}
    \end{minipage}
    \hfill
    \begin{minipage}[t]{0.495\textwidth}
        \centering
        {(b)}\\[-0.3mm]
        \includegraphics[
            width=\linewidth,
            trim=25 145 35 170,
            clip
        ]{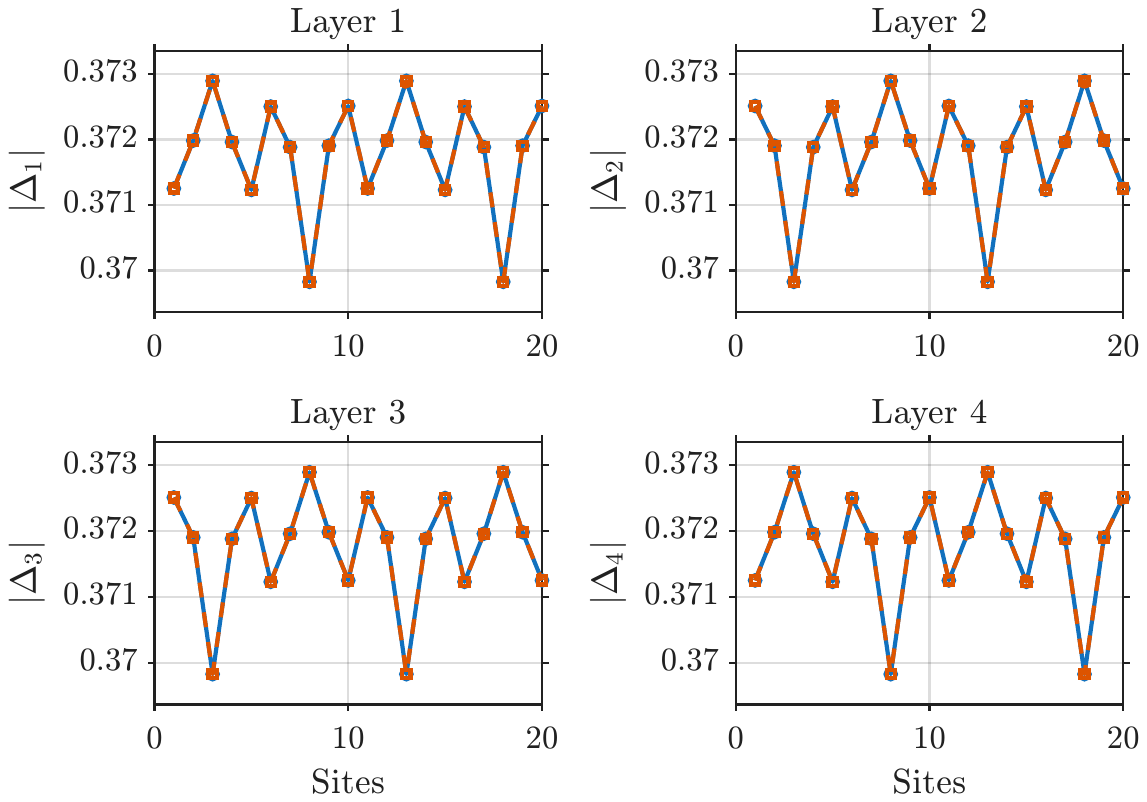}
    \end{minipage}

    \vspace{-1mm}

    \caption{
    Convergence of the representative infinite-layer solution shown in Fig.~\ref{fig:IL_vortex} with
    respect to the momentum sampling along the stacking direction,
    comparing $N_{k_z}=16$ and $N_{k_z}=20$ uniformly spaced
    $k_z$ points at $T=0$, $\Phi_p=0.025\Phi_0$, $N=80$, $t=0.25$ for a $20\times4$ MUC in the $xz$ plane.
    (a) Gauge-invariant relative phases $Q_{l,l+1}$ between
    neighboring layers. The two samplings give essentially identical
    phase profiles with the same phase-winding structure and vortex
    configuration.
    (b) Corresponding layer-resolved superconducting gap amplitudes
    $|\Delta_l|$, which are essentially indistinguishable for the two
    $k_z$ samplings.
    }
    \label{fig:kz_convergence}
\end{figure*}

\setcounter{secnumdepth}{1}

\appendix

\section{Convergence with respect to $k_z$ sampling}
\label{app:kz}

To test numerical convergence with respect to the momentum sampling along the
$\mathbf{\hat{z}}$ direction, we repeated the representative infinite-layer
calculation shown in Fig.~\ref{fig:IL_vortex} using $N_{k_z}=20$ uniformly spaced points instead of the
$N_{k_z}=16$ points used in the main calculations. The resulting
layer-resolved gap-amplitude and relative-phase profiles are essentially indistinguishable, as shown in Fig.~\ref{fig:kz_convergence}. The energies obtained with the two
samplings also agree to numerical precision. These results
indicate that $N_{k_z}=16$ is sufficient for the results presented in
the main text, in particular the phase diagram in Fig.~\ref{fig:phase_diagrams}.

\section{Anisotropy of the Josephson-vortex lattice}
\label{app:anisotropy}

As discussed in the main text, the distorted Josephson-vortex lattice of Fig.~\ref{fig:IL_vortex} can be understood in terms of the effective-mass anisotropy in the in-plane and out-of-plane directions. For an
anisotropic superconductor, the appropriate rescaling factor is
determined by the anisotropy of the superconducting stiffness~\cite{Koshelev2013}, which can be parametrized by the Cooper-pair effective masses~\cite{Iazzi2012} as
\begin{equation}
\gamma_{\mathrm{pair}}
=
\sqrt{\frac{m_z^{\mathrm{pair}}}{m_x^{\mathrm{pair}}}}
\label{eq:gamma_pair}
\end{equation}
Since the effective pair mass along direction $i$ is inversely
proportional to the corresponding pair-hopping amplitude $J_i$
\cite{Iazzi2012,Micnas1990}, for equal lattice spacings we have
\begin{equation}
\gamma_{\mathrm{pair}}
=
\sqrt{\frac{J_x}{J_z}}
\label{eq:gamma_pair_hopping}
\end{equation}

In the strong-coupling limit ($U\gg t_i$) of the attractive Hubbard model, an on-site pair moves between neighboring sites
through the virtual process
\begin{equation}
|\uparrow\downarrow,0\rangle
\longrightarrow
|\uparrow,\downarrow\rangle
\longrightarrow
|0,\uparrow\downarrow\rangle 
\end{equation}
The corresponding leading-order pair-hopping amplitude is
\cite{Micnas1990}
\begin{equation}
J_i \simeq \frac{2t_i^2}{U}
\label{eq:pair_hopping_strong}
\end{equation}
which gives
\begin{equation}
\gamma_{\mathrm{pair}}
\simeq
\frac{t_1}{t_2}
\label{eq:gamma_strong}
\end{equation}
For the parameters of Fig.~\ref{fig:IL_vortex},
 this yields
$\gamma_{\mathrm{pair}}\simeq4$ as mentioned in the main text.

Since the calculations are performed at $U=4$, the in-plane
motion is not asymptotically in the strong-coupling limit. We
therefore estimate the finite-$U$ correction using the exact
two-particle spectrum of an isolated attractive-Hubbard bond. For two neighboring sites connected by hopping $t$, the
corresponding two-site Hamiltonian is
\begin{equation}
H_{\mathrm{bond}}
=
-t\sum_{\sigma}
\left(
c_{1\sigma}^{\dagger}c_{2\sigma}
+
c_{2\sigma}^{\dagger}c_{1\sigma}
\right)
-U\sum_{i=1,2}n_{i\uparrow}n_{i\downarrow}.
\label{eq:two_site_hubbard}
\end{equation}
Here, $U>0$ is the strength of the attractive on-site interaction,
$\sigma\in\{\uparrow,\downarrow\}$ denotes the spin, and
$n_{i\sigma}=c_{i\sigma}^{\dagger}c_{i\sigma}$ is the number operator
for spin $\sigma$ on site $i=1,2$.
Consider the doublon states
\begin{equation}
|D_1\rangle=|\uparrow\downarrow,0\rangle,
\qquad
|D_2\rangle=|0,\uparrow\downarrow\rangle
\end{equation}
and the singly occupied spin-singlet state
\begin{equation}
|S\rangle
=
\frac{1}{\sqrt{2}}
\left(
|\uparrow,\downarrow\rangle
-
|\downarrow,\uparrow\rangle
\right)
\end{equation}
Introducing
\begin{equation}
|D_{\pm}\rangle
=
\frac{|D_1\rangle\pm|D_2\rangle}{\sqrt{2}}
\end{equation}
the antisymmetric state $|D_-\rangle$ decouples from $|S\rangle$
and has energy
\begin{equation}
E_-=-U.
\end{equation}
The remaining $\{|D_+\rangle,|S\rangle\}$ sector is described by
\begin{equation}
H_{\mathrm{red}}
=
\begin{pmatrix}
-U & -2t\\
-2t & 0
\end{pmatrix}
\end{equation}
whose lower eigenvalue is
\begin{equation}
E_{\mathrm{low}}
=
\frac{-U-\sqrt{U^2+16t^2}}{2}
\end{equation}

An effective two-state Hamiltonian in the doublon subspace can be
written as
\begin{equation}
H_{\mathrm{pair}}
=
\epsilon\sum_{i=1,2}|D_i\rangle\langle D_i|
-
J_{\mathrm{pair}}
\left(
|D_1\rangle\langle D_2|
+
|D_2\rangle\langle D_1|
\right)
\end{equation}
Its symmetric and antisymmetric states are separated in energy by
$2J_{\mathrm{pair}}$. Matching this splitting to the two lowest
levels of the exact two-site spectrum gives
\begin{equation}
J_{\mathrm{pair}}(U,t)
=
\frac{\sqrt{U^2+16t^2}-U}{4}
\label{eq:pair_hopping_finiteU}
\end{equation}
For $t/U\ll1$,
\begin{equation}
J_{\mathrm{pair}}(U,t)
=
\frac{2t^2}{U}
+
\mathcal{O}\left(\frac{t^4}{U^3}\right)
\end{equation}
recovering Eq.~\eqref{eq:pair_hopping_strong}. We emphasize that
Eq.~\eqref{eq:pair_hopping_finiteU} is exact for an isolated
two-site bond and is used here only to estimate the finite-$U$
correction to the pair anisotropy.

Applying Eq.~\eqref{eq:pair_hopping_finiteU} separately to the
in-plane and interlayer hoppings gives
\begin{equation}
\gamma_{\mathrm{pair}}
\simeq
\left[
\frac{\sqrt{U^2+16t_1^2}-U}
     {\sqrt{U^2+16t_2^2}-U}
\right]^{1/2}
\label{eq:gamma_finiteU}
\end{equation}
For $U=4$, $t_1=1$, and $t_2=0.25$, we obtain
\begin{equation}
\gamma_{\mathrm{pair}}\simeq3.7,
\end{equation}
close to the strong-coupling estimate. We therefore take
$\gamma_{\mathrm{pair}}\simeq3.7$--$4$ as a physically motivated
range for the parameters of Fig.~\ref{fig:IL_vortex}. Rescaling the stacking direction using
$\gamma_{\mathrm{pair}}\simeq3.7$--$4$ gives a characteristic
vortex-lattice angle of approximately $56^\circ$--$58^\circ$,
close to the $60^\circ$ angle of an equilateral triangular
lattice as mentioned in the main text.

\section{Suppression of the BCS phase with increasing layer number}
\label{app:bcs_suppression}

Here we make explicit the layer-number dependence of the
phenomenological GL argument discussed in the main text, following
Ref.~\cite{Yan2026}. For an in-plane magnetic field $\mathbf{B}=B\hat{\mathbf y}$, we choose the gauge
$\mathbf{A}=Bz\,\hat{\mathbf x}$. A layer at position $z_l$ then
prefers a phase gradient

\begin{equation}
q_l=\frac{2eBz_l}{\hbar}
\end{equation}

The orbital-FFLO state can accommodate different preferred
phase gradients in different layers. In contrast, constraining all layers
to have a common phase gradient $q$ gives an additional orbital
energy cost
\begin{equation}
\Delta E_{\mathrm{orb}}
\propto
\sum_l (q-q_l)^2 
\label{eq:orbital_cost}
\end{equation}
Minimizing this expression with respect to the common $q$
gives
$q=\bar{q}=P^{-1}\sum_l q_l$.
The corresponding orbital energy cost per layer is therefore
\begin{equation}
\frac{\Delta E_{\mathrm{orb}}}{P}
\propto
\left(\frac{2eB}{\hbar}\right)^2
\frac{1}{P}
\sum_l (z_l-\bar{z})^2 
\label{eq:orbital_cost_per_layer}
\end{equation}
where $\bar{z}=\frac{1}{P}\sum_l z_l$. For the inversion-symmetric stack considered here, $\bar{z}=0$,
and hence the optimal common phase gradient is $\bar{q}=0$. 
For $P$ equally spaced layers with interlayer spacing $d$,
\begin{equation}
\frac{1}{P}
\sum_{l=1}^{P}(z_l-\bar{z})^2
=
\frac{d^2(P^2-1)}{12}
\end{equation}
so that
\begin{equation}
\frac{\Delta E_{\mathrm{orb}}}{P}
\propto
\left(\frac{2eBd}{\hbar}\right)^2
\frac{P^2-1}{12}
\label{eq:orbital_cost_scaling}
\end{equation}
Thus, the orbital energy cost per layer of the uniform BCS state
grows as $P^2$, providing the layer-number dependence used in the
discussion of Fig.~\ref{fig:phase_diagrams}(a)--~\ref{fig:phase_diagrams}(c).

\section{Local supercurrents and boundary twists}
\label{app:currents}
\begin{figure*}[t]
    \centering


    \begin{minipage}[t]{0.38\textwidth}
        \makebox[\linewidth][c]{{(a)}}
        \centering
        \includegraphics[width=\linewidth]
        {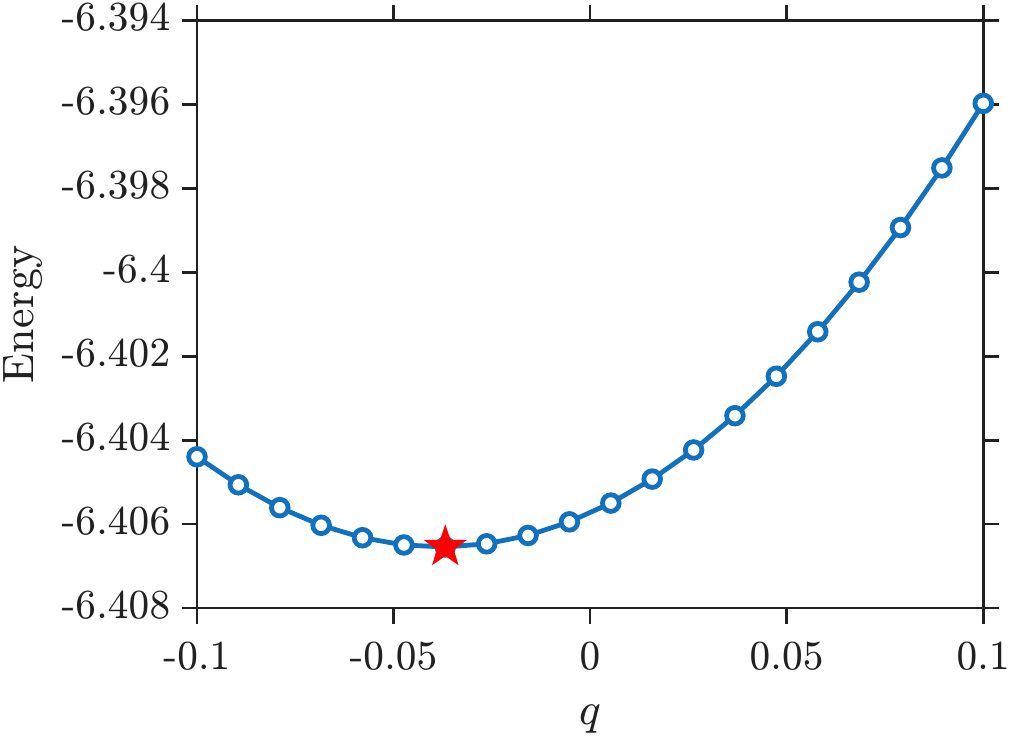}
    \end{minipage}
    \hspace{0.06\textwidth}
    \begin{minipage}[t]{0.38\textwidth}
        \makebox[\linewidth][c]{{(b)}}
        \centering
        \includegraphics[width=\linewidth]
        {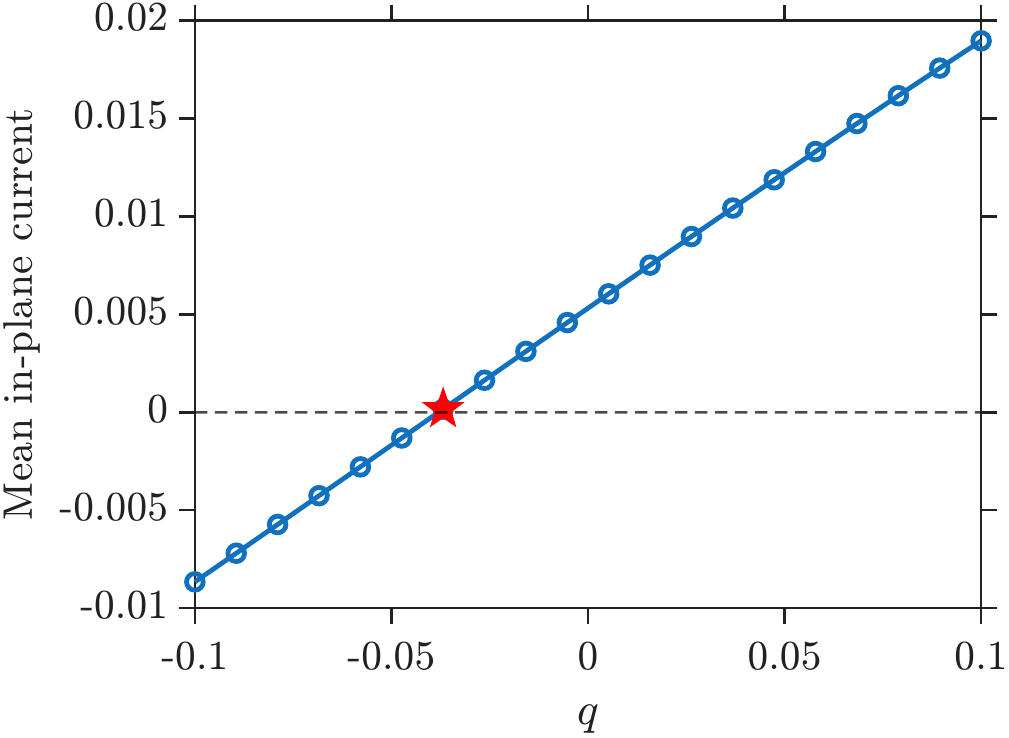}
    \end{minipage}

    \vspace{0.3cm}


    \begin{minipage}[t]{0.48\textwidth}
    \makebox[\linewidth][c]{(c)}\\[1.5mm]
    \centering
    \includegraphics[
        width=\linewidth,
        height=6.3cm
    ]{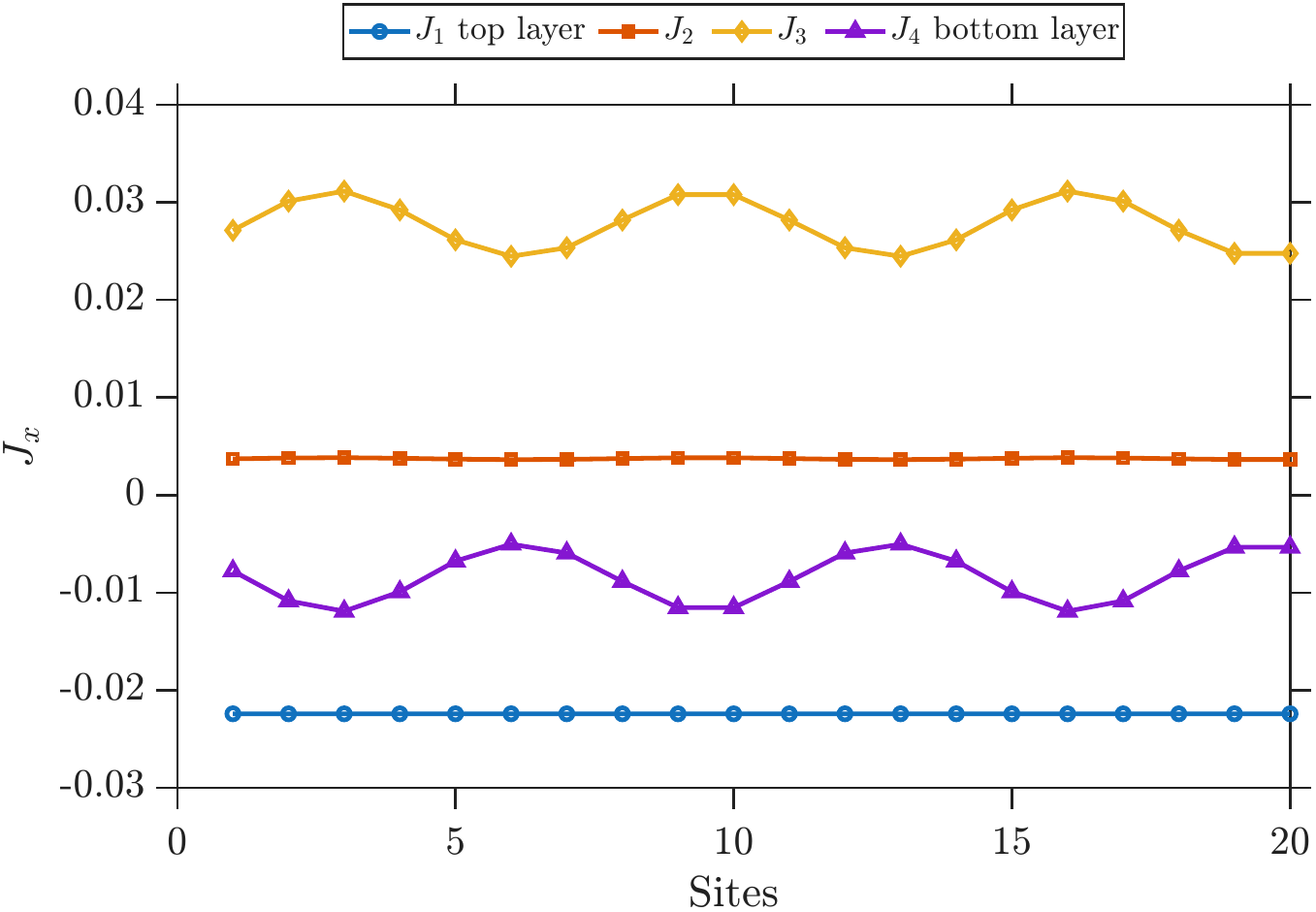}
\end{minipage}
    \hfill
    \begin{minipage}[t]{0.48\textwidth}
    \makebox[\linewidth][c]{(d)}\\[1.5mm]
    \centering
    \includegraphics[
        width=\linewidth,
        height=6.3cm
    ]{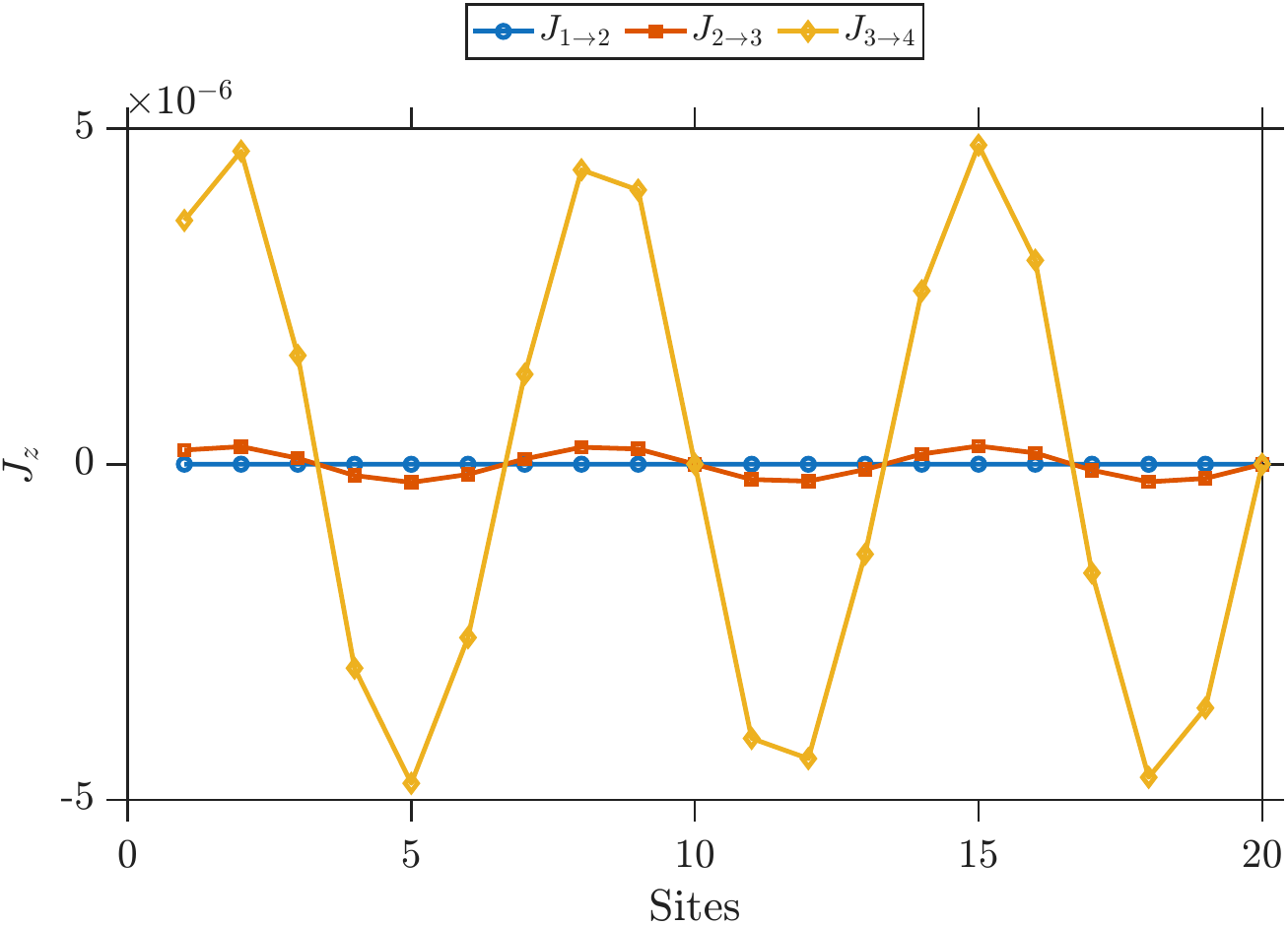}
\end{minipage}

  \caption{
Boundary-twist minimization and local supercurrents for
the $t=0.20$ surface-localized vortex state shown in
Fig.~\ref{fig:4layer_rearr}(a), with
$\Phi_p=0.030\Phi_0$, $N=20$, $U=4$, $\mu=-3$, and $T=0$.
(a) Energy per lattice site as a function of the boundary twist $q$.
The red star marks the minimum at $q\simeq-0.03684$.
(b) Mean in-plane current as a function of $q$. The current
vanishes at essentially the same twist, consistent with the
relation between the free-energy derivative and the total current
(Eq.~\eqref{eq:twist}).
(c) Layer-resolved local intralayer currents $J_x^l(j)$
evaluated at $q=-0.03684$.
(d) Corresponding interlayer currents
$J_z^{l\rightarrow l+1}(j)$. The three spatial modulations of
$J_z^{3\rightarrow4}$ correspond to the three vortices in the
$3$--$4$ interlayer junction.
}
    \label{fig:currents_twist}
\end{figure*}
As discussed in the main text, a uniform boundary twist $q$ is
introduced along $\hat{\mathbf{x}}$ to eliminate the finite-size
net equilibrium current. 
To compute the local supercurrents in a finite P-layer system, we consider the intralayer and interlayer current operators 
\begin{equation}
\hat{J}^{\,l}_{\hat{\delta},\mathbf{j}}
=
-\frac{\partial H}{\partial
A^{l}_{\hat{\delta},\mathbf{j}}},
\qquad
\hat{J}^{\,l}_{z,\mathbf{j}}
=
-\frac{\partial H}{\partial
A^{l}_{z,\mathbf{j}}}
\end{equation}
where
$\hat{\delta}\in\{\hat{\mathbf{x}},\hat{\mathbf{y}}\}$.
For the Hamiltonian in Eq.~\eqref{eq:hamiltonian}, the intralayer and interlayer bond currents are, respectively,
\begin{align}
\langle\hat{J}^{l}_{\hat{\delta},\mathbf{j}}\rangle
&=
it_{1}
\sum_{\sigma}
\Big[
e^{iA^{l}_{\hat{\delta},\mathbf{j}}}
\left\langle
c^{\dagger}_{l,\mathbf{j}+\hat{\delta},\sigma}
c_{l,\mathbf{j},\sigma}
\right\rangle
\nonumber\\
&\qquad\qquad
-
e^{-iA^{l}_{\hat{\delta},\mathbf{j}}}
\left\langle
c^{\dagger}_{l,\mathbf{j},\sigma}
c_{l,\mathbf{j}+\hat{\delta},\sigma}
\right\rangle
\Big],
\label{eq:intralayer_current}
\\
\langle\hat{J}^{l}_{z,\mathbf{j}}\rangle
&=
it
\sum_{\sigma}
\Big[
e^{iA^{l}_{z,\mathbf{j}}}
\left\langle
c^{\dagger}_{l+1,\mathbf{j},\sigma}
c_{l,\mathbf{j},\sigma}
\right\rangle
\nonumber\\
&\qquad\qquad
-
e^{-iA^{l}_{z,\mathbf{j}}}
\left\langle
c^{\dagger}_{l,\mathbf{j},\sigma}
c_{l+1,\mathbf{j},\sigma}
\right\rangle
\Big]
\label{eq:interlayer_current}
\end{align}
Note that for the solutions considered here, $\langle\hat{J}^{l}_{\hat{y},\mathbf{j}}\rangle$ vanishes.

Fig.~\ref{fig:currents_twist} illustrates the role of the boundary
twist and the resulting local-current structure for the $t=0.20$
surface-localized vortex state shown in Fig.~\ref{fig:4layer_rearr}(a). The energy
per site exhibits a minimum at $q\simeq-0.03684$, while the mean
in-plane current vanishes at essentially the same value of $q$, as
expected from the relation between the free-energy derivative and
the total current discussed in the main text (see Eq.~\eqref{eq:twist}). The
layer-resolved intralayer currents remain finite and spatially
varying, while the three spatial modulations of
$J_z^{3\rightarrow4}$ correspond to the three vortices in the
$3$--$4$ interlayer junction. Thus, the boundary-twist
minimization removes the spurious finite-size net current while
retaining the nonuniform equilibrium-current pattern associated
with the Josephson-vortex configuration.

\section{Additional four-layer calculations}
\label{app:add_4L}

\begin{figure*}[t]
    \centering

\begin{minipage}[t]{0.32\textwidth}
    \centering
    \includegraphics[width=\linewidth]{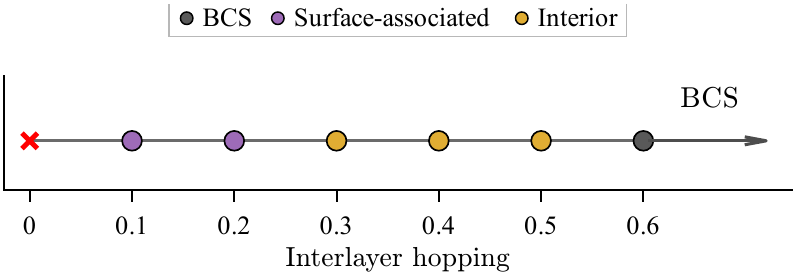}\\[-0.01mm]
    \makebox[\linewidth][l]{\hspace{2.6cm}(a)}
\end{minipage}%
\hspace{0.01\textwidth}%
\begin{minipage}[t]{0.32\textwidth}
    \centering
    \includegraphics[width=\linewidth]{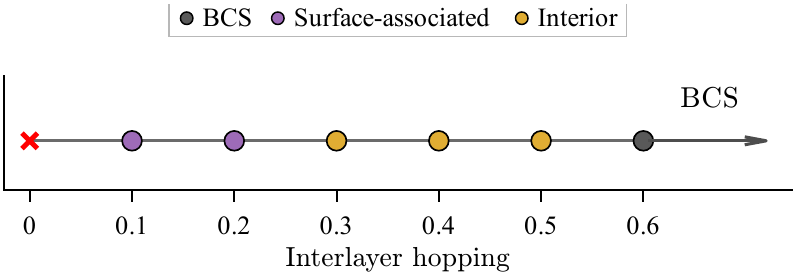}\\[-0.01mm]
    \makebox[\linewidth][l]{\hspace{2.6cm}(b)}
\end{minipage}%
\hspace{0.01\textwidth}%
\begin{minipage}[t]{0.32\textwidth}
    \centering
    \includegraphics[width=\linewidth]{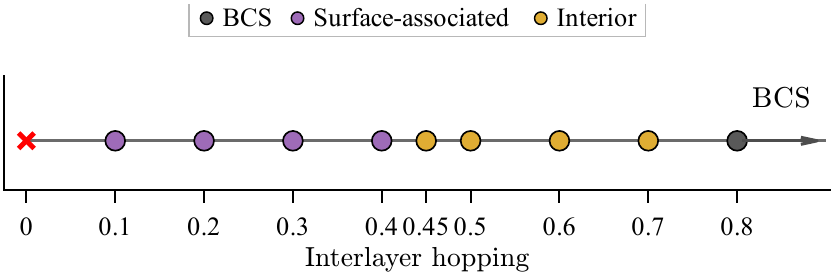}\\[-0.01mm]
    \makebox[\linewidth][l]{\hspace{2.6cm}(c)}
\end{minipage}

\vspace{0.18cm}

   \makebox[\textwidth][c]{%
    \begin{minipage}[t]{0.35\textwidth}
        \centering
        \includegraphics[width=\linewidth,height=6.6cm]{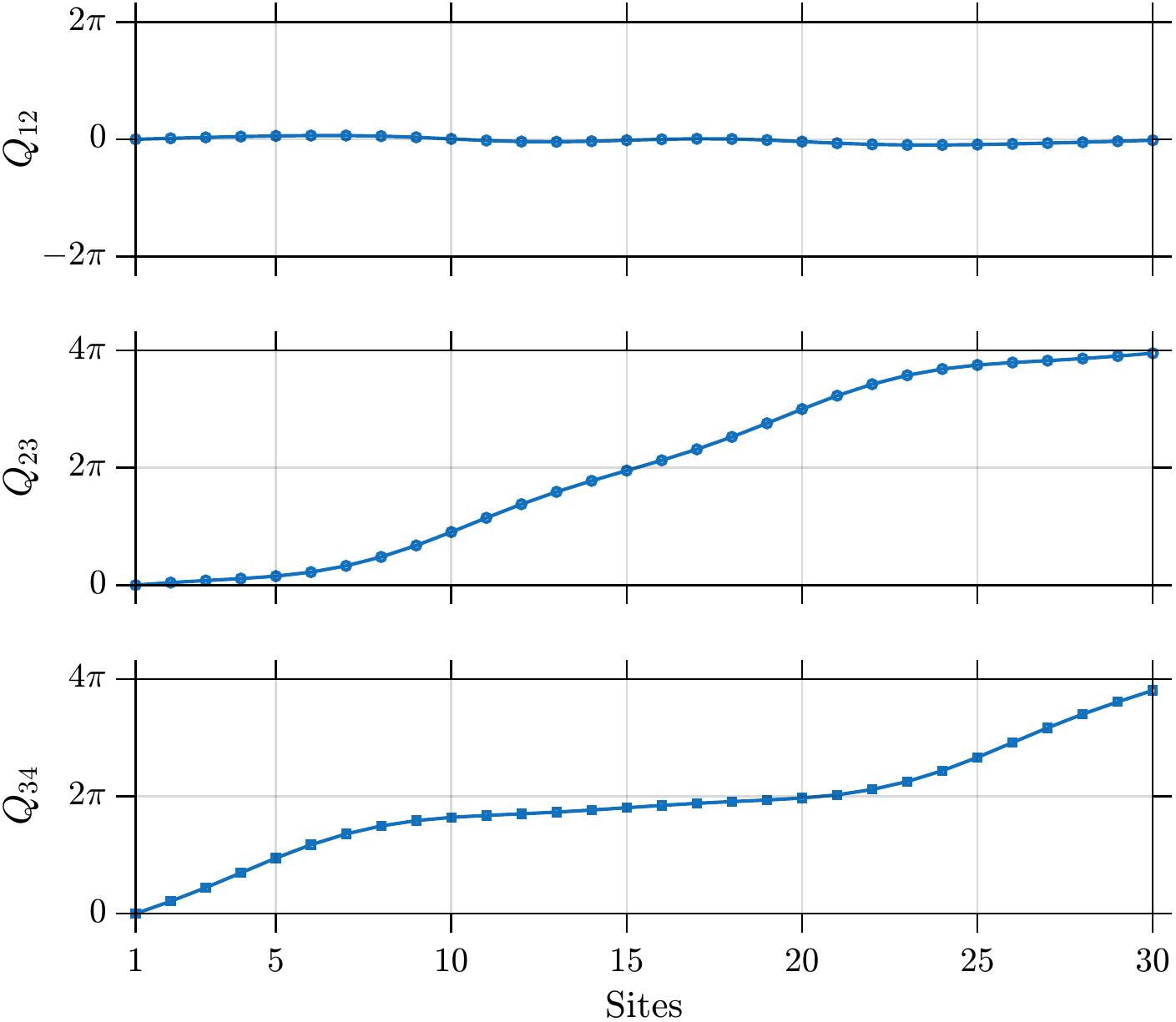}\\[-0.01mm]
    \makebox[\linewidth][l]{\hspace{3.2cm}(d)}
    \end{minipage}
    \hspace{0.02\textwidth}
    \begin{minipage}[t]{0.35\textwidth}
        \centering
        \includegraphics[width=\linewidth,height=6.6cm]{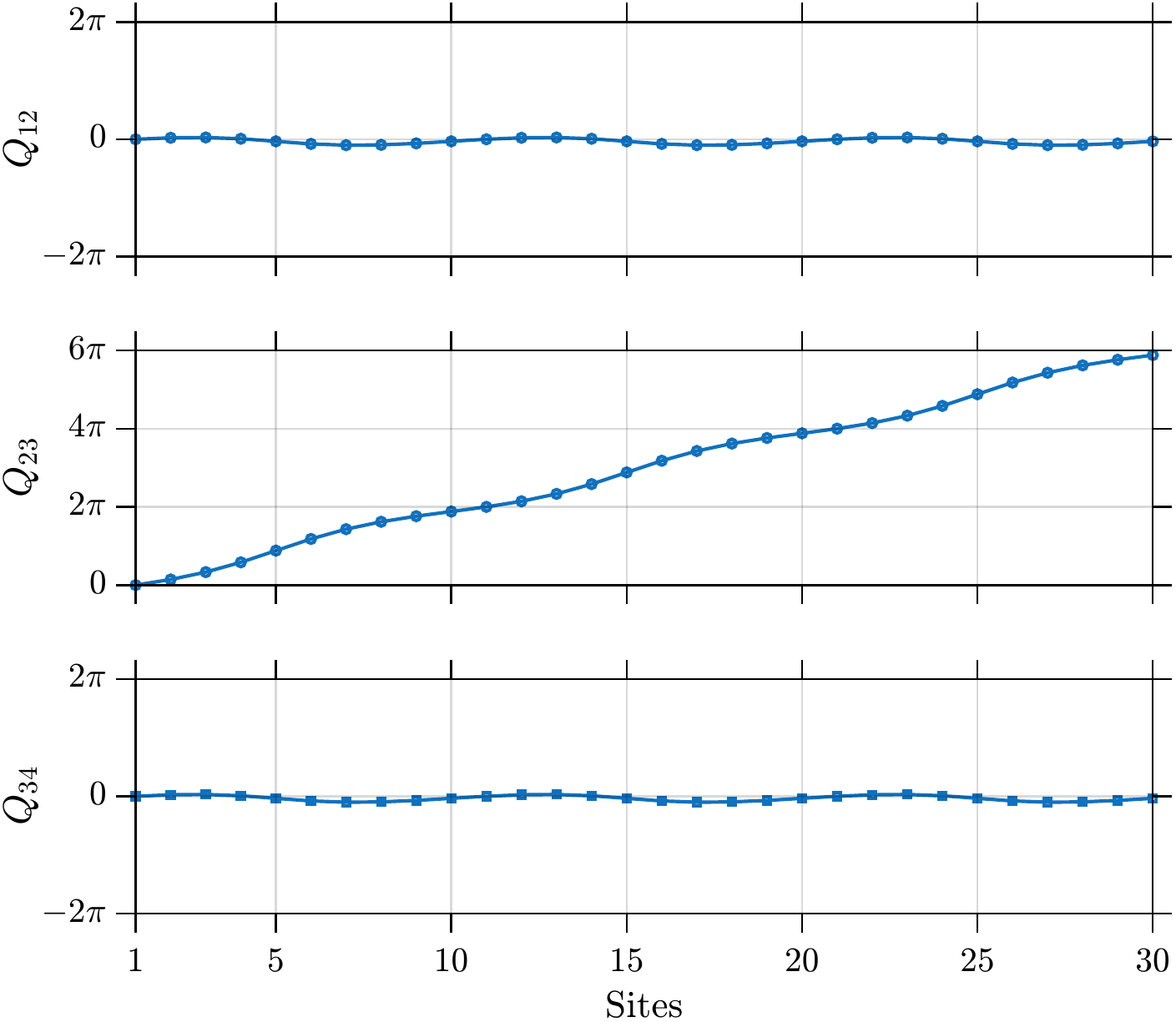}\\[-0.01mm]
    \makebox[\linewidth][l]{\hspace{3.2cm}(e)}
    \end{minipage}
}

    \vspace{0.2cm}

    \caption{
Additional four-layer calculations demonstrating the persistence of the
interlayer-hopping-driven vortex rearrangement across different
system parameters. Figures (a)--(c) show the evolution of the ground state
configuration as a function of the
interlayer hopping $t$. Figures (a) and (b) correspond to
$\Phi_p=0.028\Phi_0$ for system sizes $N=20$ and $N=30$, respectively,
while panel (c) corresponds to $\Phi_p=0.035\Phi_0$ and $N=20$.
The configurations are classified as BCS, surface-associated or interior
according to the superconducting state and
the location of the Josephson vortices. The $t=0$ point is left unclassified
because the layers are completely decoupled in this limit. Panels (d) and (e) show
representative gauge-invariant relative-phase profiles
$Q_{12}$, $Q_{23}$, and $Q_{34}$ from the $N=30$ calculation in panel (b)
at $t_2=0.20$ and $t_2=0.30$, respectively. At $t_2=0.20$, the vortices form a staggered configuration, with two
vortices in the central $2$--$3$ junction and two in an outer junction,
whereas at $t_2=0.30$ all three vortices are concentrated in the central
junction, illustrating the rearrangement toward the interior. The numerical parameters used are
$\mu=-3.0$, $U=4$, and $T=0$.
}
\label{fig:additional}
    
\end{figure*}

Here we present additional self-consistent four-layer calculations to
test the robustness of the interlayer-hopping-driven Josephson-vortex
rearrangement discussed in the main text. We examine the evolution of
the lowest-energy vortex configurations as a function of the interlayer
hopping $t$ for different magnetic fluxes and system sizes. The
Hamiltonian parameters are fixed at $\mu=-3.0$, $U=4$, and $T=0$.

Figure~\ref{fig:additional}(a) shows the evolution at
$\Phi_p=0.028\Phi_0$ for $N=20$. With increasing $t$, the preferred
configuration evolves from surface-associated vortices to vortices
localized in the interior of the stack, before the system eventually enters the BCS state. To test whether this behavior
is tied to the finite system size, Fig.~\ref{fig:additional}(b) shows
the corresponding calculation at the same magnetic flux for the larger
system $N=30$. The same sequence of vortex reorganization is recovered,
with increasing interlayer hopping systematically favoring configurations
in which the vortices are concentrated in the interior interlayer
junction. Figure~\ref{fig:additional}(c) provides a further check at the
larger magnetic flux $\Phi_p=0.035\Phi_0$ for $N=20$, where the same
tendency toward interior vortex configurations is again found.

The microscopic character of this rearrangement is illustrated by the
representative gauge-invariant relative-phase profiles in
Figs.~\ref{fig:additional}(d) and \ref{fig:additional}(e), taken from
the $N=30$, $\Phi_p=0.028\Phi_0$ calculation. At $t=0.20$, the system
forms a staggered vortex configuration, with two vortices in the central
$2$--$3$ interlayer junction and two in an outer junction. Increasing
the interlayer hopping to $t=0.30$ reorganizes the phase-winding pattern
so that all three vortices are concentrated in the central $2$--$3$
junction. Thus, the additional calculations consistently reproduce the
same interlayer-hopping-driven migration of vortices toward the interior
under changes in both system size and magnetic flux, demonstrating that
the rearrangement is a robust feature of the four-layer solutions over
the parameter regimes examined here.

\begin{figure*}[t]
    \centering

    \includegraphics[width=0.42\textwidth, height=6.8cm]{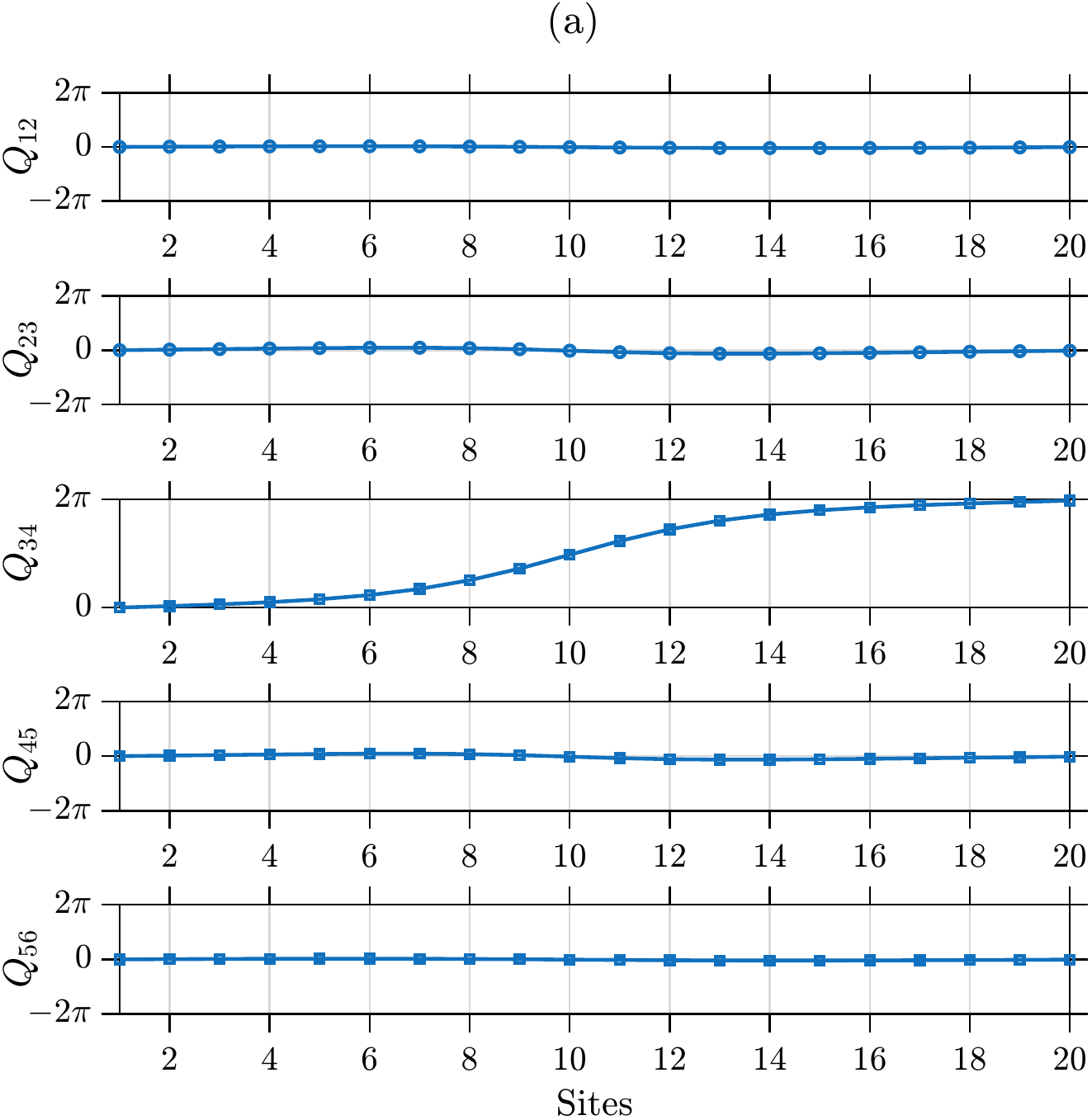}%
    \hspace{0.03\textwidth}
    \includegraphics[width=0.42\textwidth, height=6.8cm]{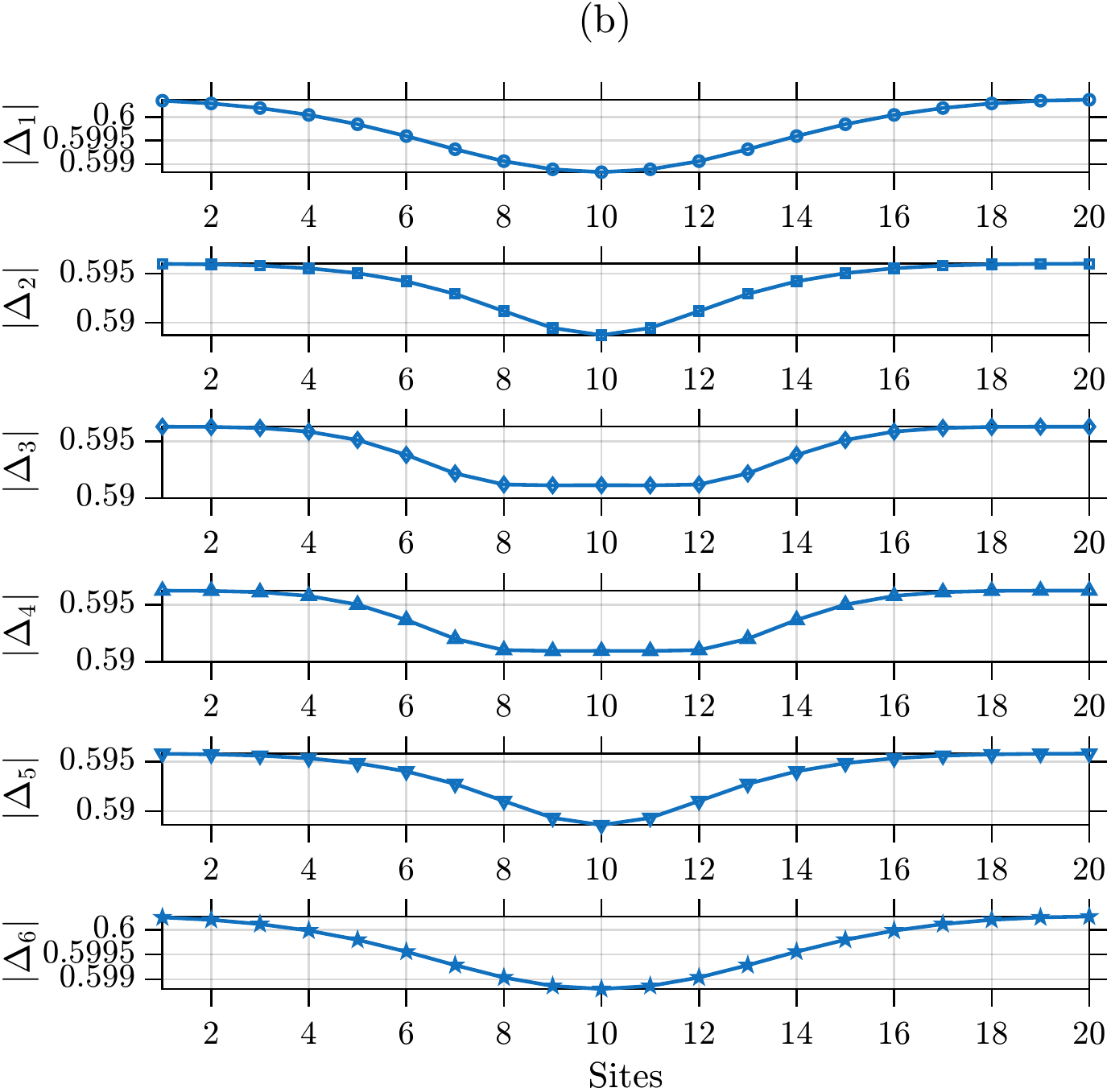}

    \caption{
Six-layer interior vortex configuration at $T=0$, $\Phi_p=0.015\Phi_0$,
$N=20$, and $t=0.30$, with optimized boundary twist
$q=-0.07854$. Panel (a) shows the gauge-invariant relative phases
$Q_{l,l+1}$ between neighboring layers. The phase winding is
concentrated in the central $3$--$4$ interlayer junction, identifying
a Josephson vortex located deep within the six-layer stack.
Panel (b) shows the corresponding layer-resolved gap amplitudes
$|\Delta_l|$. The central vortex produces a broad suppression of the
order parameter throughout the interior layers, with local gap
depressions of approximately $0.9\%$, while the suppression on the
two surface layers is only about $0.25\%$. Thus, although the vortex
resides in the central junction, it leaves a finite surface-layer
modulation (although weaker than in the representative four-layer examples), providing a qualitative extension of the surface-imprint
result to a thicker finite stack.
}
\label{fig:6layer}
\end{figure*}

\section{Six-layer calculations}
\label{app:6L}

To examine whether an interior Josephson vortex can leave a surface
signature in a thicker finite stack, we performed additional
self-consistent calculations for a six-layer system at $T=0$,
$\Phi_p=0.015\Phi_0$, and $N=20$. For this qualitative extension,
the initial guesses were restricted to configurations containing at
most four seeded vortices, as appropriate for the chosen value of
$\Phi_p$ and $N$. For the representative interior-vortex configuration
discussed below, we additionally optimized the boundary twist $q$.

As shown in Fig.~\ref{fig:6layer}, for $t_2=0.30$ the phase winding
is concentrated in the central $3$--$4$ interlayer junction, placing
the Josephson vortex deep within the six-layer stack. The corresponding
gap profiles show a broad suppression of the superconducting order
parameter throughout the interior layers, of order $0.9\%$, while the
surface layers are suppressed by only about $0.25\%$. Thus, despite
being located in the central junction, the vortex produces a finite
modulation of the surface-layer gap. The resulting surface imprint is
substantially weaker than in the representative four-layer case shown
in Fig.~\ref{fig:4layer_rearr}, providing a qualitative extension of
the surface-signature result to a thicker finite stack.

\section{PBC calculations}
\label{app:bc}

\begin{figure*}[t]
    \centering

    \begin{minipage}[t]{0.49\textwidth}
        \centering
        \vspace{0pt}
        {\large{(a)}}\\[2mm]
        \includegraphics[height=6.5cm,keepaspectratio]
        {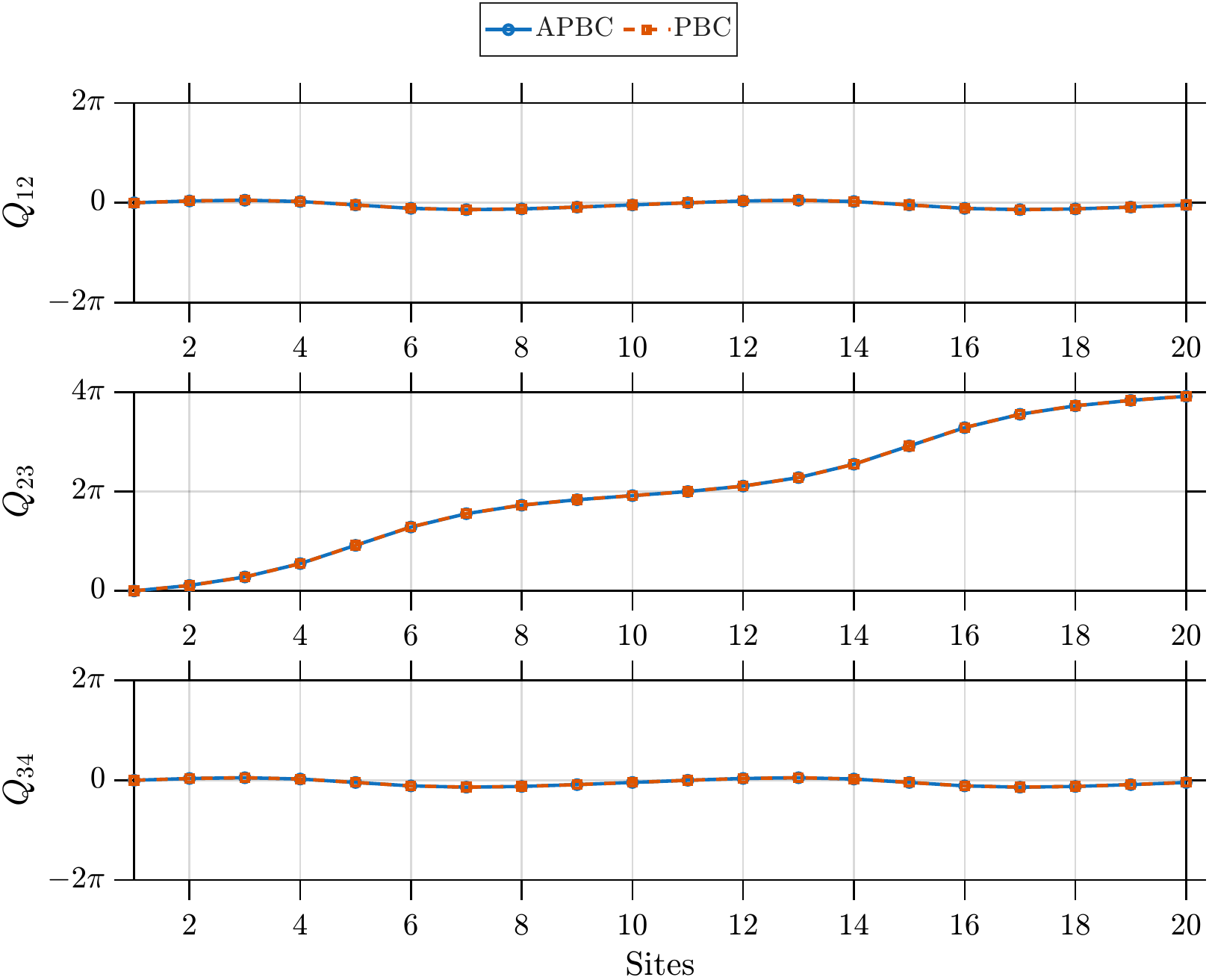}
    \end{minipage}
    \hfill
    \begin{minipage}[t]{0.49\textwidth}
        \centering
        \vspace{0pt}
        {\large{(b)}}\\[2mm]
        \includegraphics[height=6.5cm,keepaspectratio]
        {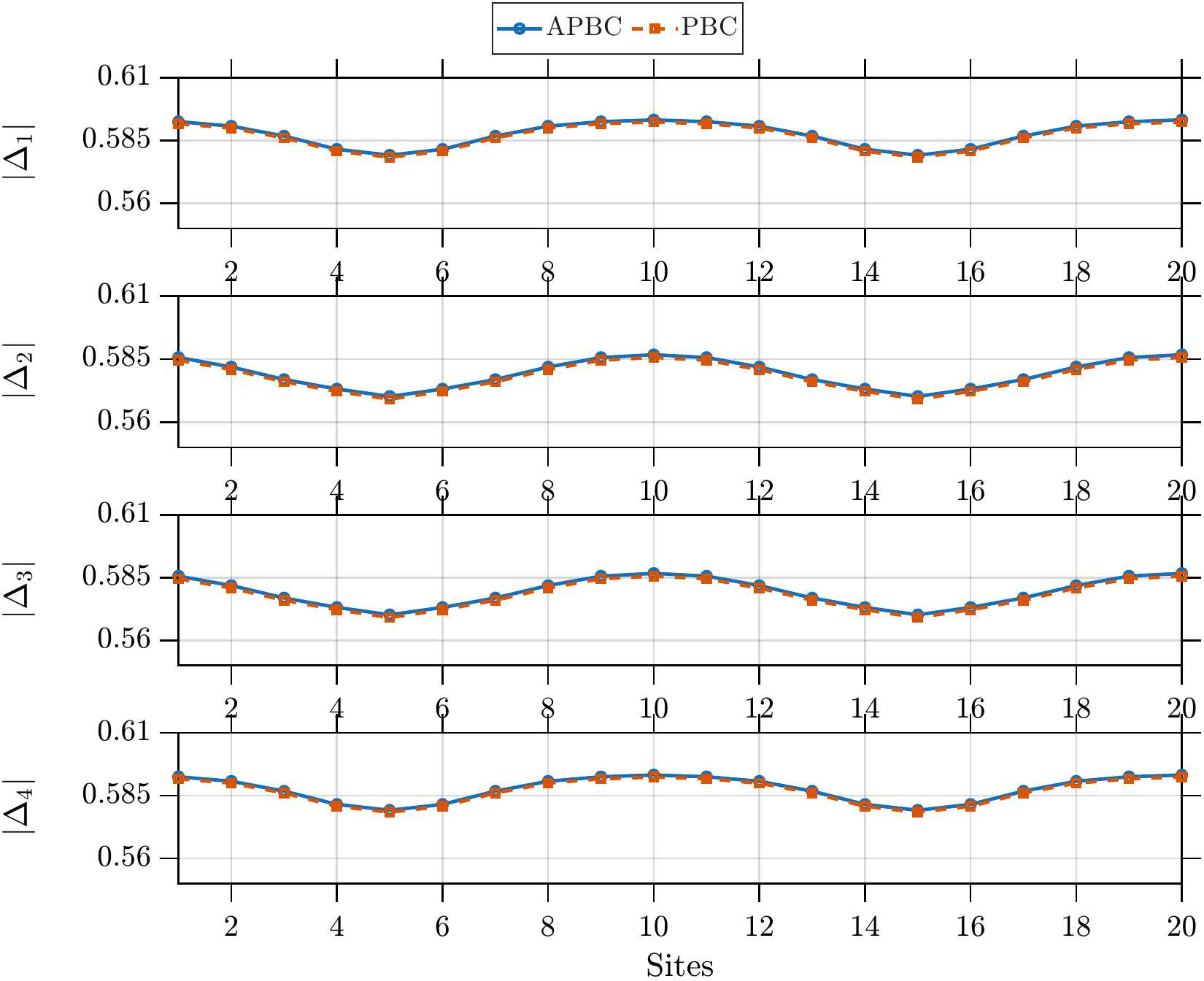}
    \end{minipage}

    \caption{
    Comparison of the self-consistent solution obtained using
    APBCs and PBCs along the $y$ direction for
    $\Phi_p=0.030\Phi_0$, $t=0.40$, and $N=20$ for the state shown
    in Fig.~\ref{fig:4layer_rearr}(b).
    (a) Gauge-invariant relative phases
    $Q_{12}$, $Q_{23}$, and $Q_{34}$.
    (b) Layer-resolved gap amplitudes $|\Delta_l|$.
    The two boundary conditions yield nearly identical profiles and
    the same vortex configuration.
    }
    \label{fig:BC_robustness}
\end{figure*}
\begin{figure*}[t]
    \centering

    \begin{minipage}[t]{0.46\textwidth}
        \centering
        {\large (a)}\\[-0.01mm]
        \includegraphics[width=\linewidth]
        {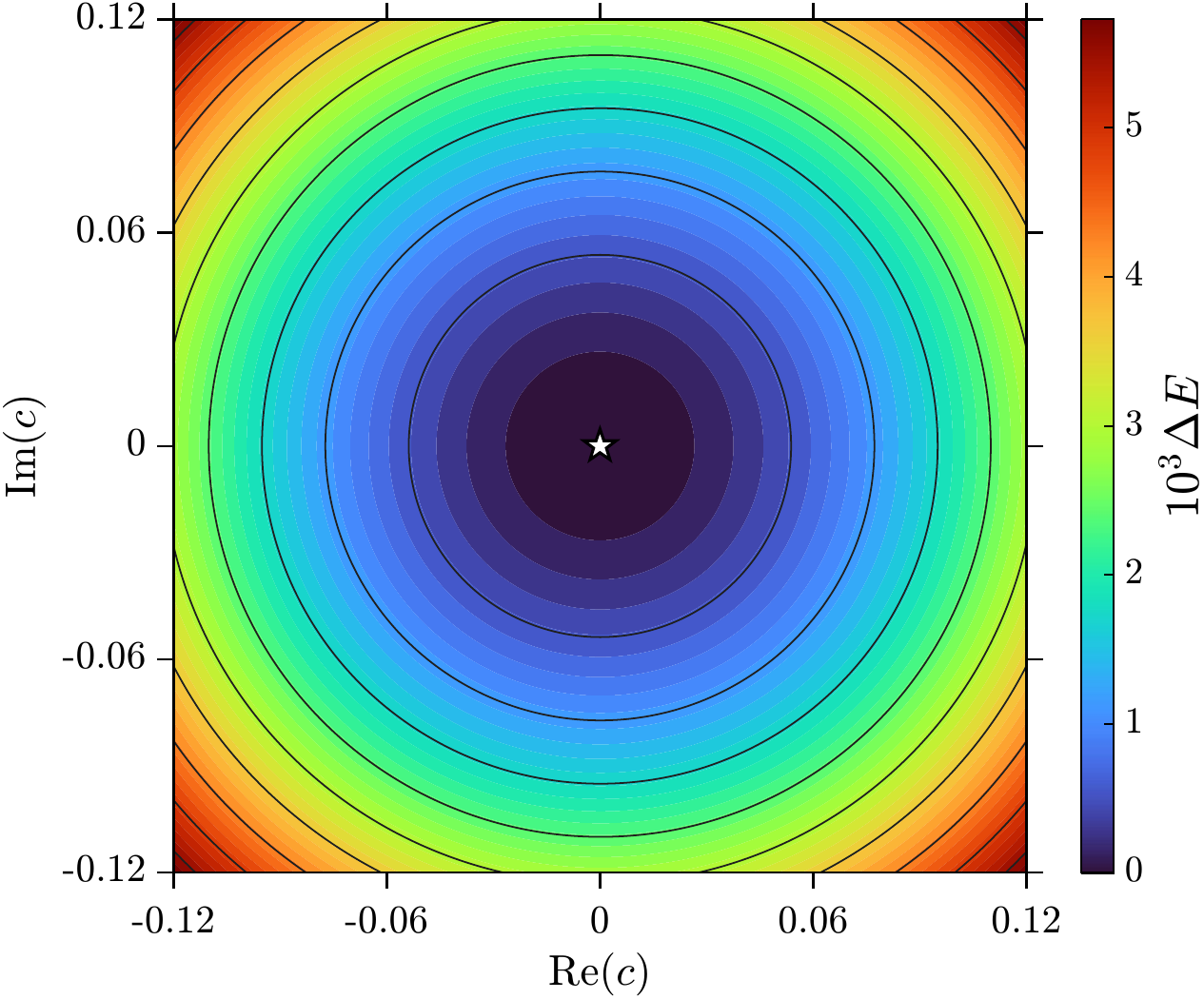}
    \end{minipage}
    \hspace{0.03\textwidth}
    \begin{minipage}[t]{0.46\textwidth}
        \centering
        {\large (b)}\\[-.01mm]
        \includegraphics[width=\linewidth]
        {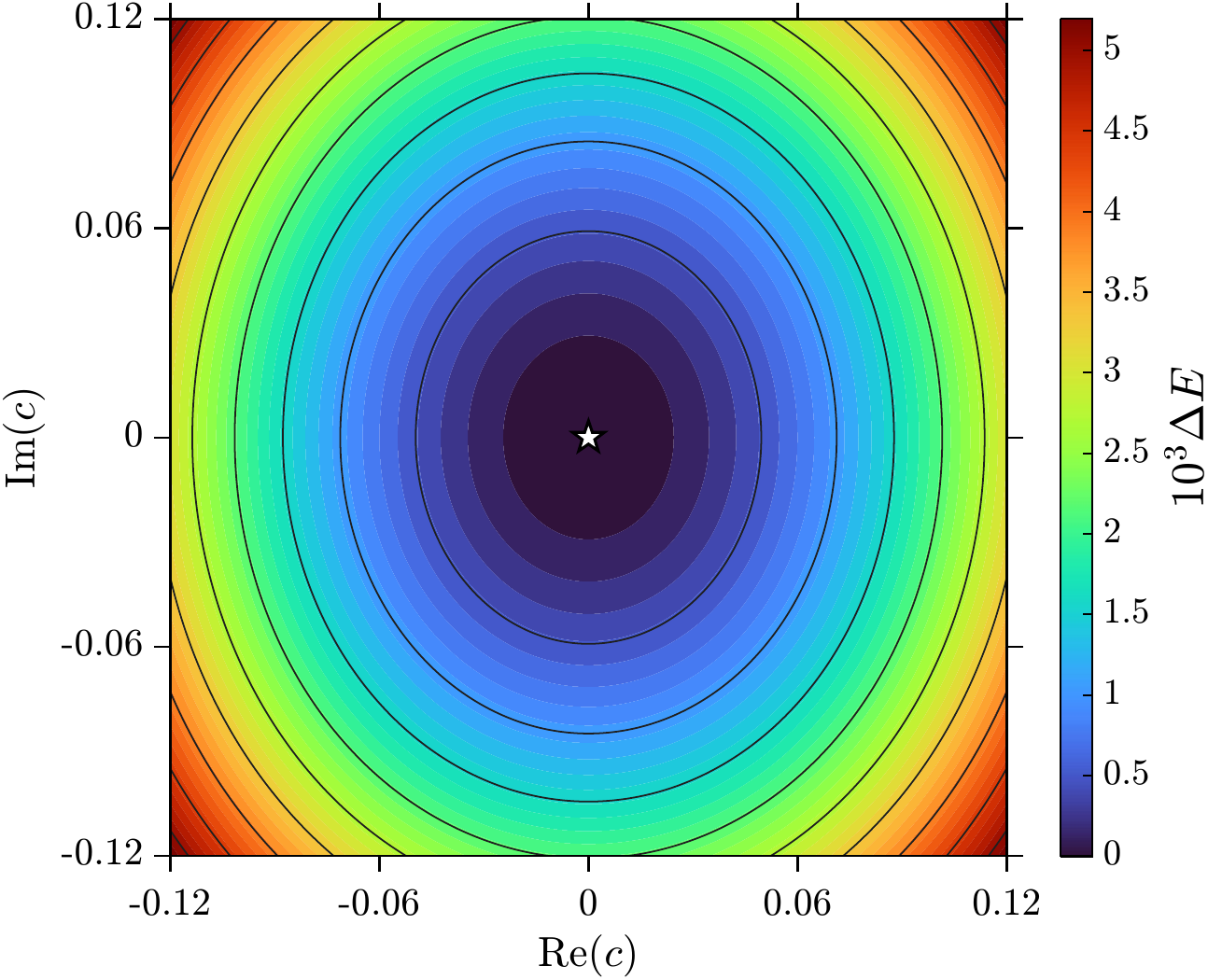}
    \end{minipage}

    \caption{
    Energy landscapes under complex perturbations
    $c=c_{\mathrm R}+ic_{\mathrm I}$ of the self-consistent four-layer
    solutions corresponding to Figs.~\ref{fig:4layer_rearr}(a) and
    \ref{fig:4layer_rearr}(b), for (a) $t=0.20$ and (b) $t=0.40$,
    respectively. The color scale shows
    $10^{3}[E(c)-E(0)]$. In both cases, the minimum occurs at $c=0$,
    indicating local stability against the class of perturbations
    considered. The smooth color maps are obtained by interpolation of
    the calculated $9\times9$ grids. The white star marks the unperturbed self-consistent solution at $c=0$.
    }
    \label{fig:stability_landscape}
\end{figure*}

As an additional check of the sensitivity of the finite-size solutions to
the choice of boundary conditions, we repeated the representative four-layer
calculations shown in Fig.~\ref{fig:4layer_rearr} after changing the boundary condition along the $y$ direction
from antiperiodic to periodic. Figure~\ref{fig:BC_robustness} shows this comparison for the
bulk-vortex state shown in Fig.~\ref{fig:4layer_rearr}(b), at $\Phi_p=0.030\Phi_0$,
$t=0.40$, and $N=20$. Both the gauge-invariant relative-phase profiles
$Q_{12}$, $Q_{23}$, and $Q_{34}$ and the layer-resolved gap amplitudes
$|\Delta_l|$ are nearly indistinguishable for the two choices of boundary
condition, with the vortex positions remaining unchanged. We similarly find that the vortex locations remain unchanged for the other configuration shown in Fig.~\ref{fig:4layer_rearr}(a). We have also verified that the infinite-layer state shown in Fig.~\ref{fig:IL_vortex} remains unchanged under PBCs.

\section{Local stability analysis}
\label{app:stability}

To further test the stability of the self-consistent four-layer solutions,
we perturb the converged order parameter according to
\begin{equation}
    \Delta_{l,j} \rightarrow \Delta_{l,j}+c,
    \qquad
    c=c_{\mathrm R}+ic_{\mathrm I},
\end{equation}
and evaluate the corresponding energy directly, without performing any
further self-consistent iterations. We compute
$\Delta E(c)=E(c)-E(0)$ on a $9\times9$ grid in the complex-$c$ plane.
For both representative configurations in
Figs.~\ref{fig:4layer_rearr}(a) and \ref{fig:4layer_rearr}(b), the
energy landscape has a unique minimum at $c=0$ and increases smoothly
away from the unperturbed solution
[Fig.~\ref{fig:stability_landscape}]. This calculation probes a
2D slice of the full order-parameter fluctuation space and
therefore provides a local-stability check only within this class of
perturbations. A complete Hessian analysis with respect to all independent
order-parameter fluctuations is beyond the scope of the present work.

\bibliographystyle{unsrtnat}
\bibliography{mahata}

\end{document}